\documentclass[12pt,onecolumn,letterpaper]{IEEEtran}
\usepackage[T1]{fontenc}
\usepackage[latin1]{inputenc}
\usepackage{color}
\usepackage{psfrag}
\usepackage[dvips]{graphicx}
\usepackage{tikz}
\usepackage{pgfplots}
\pgfplotsset{compat=1.12}
\usepackage{enumerate}
\usepackage{rotating}
\usepackage{srcltx}
\usepackage{booktabs}
\usepackage{multicol}
\usepackage{enumitem}
\usepackage{url}
\usepackage[noadjust]{cite}
\usepackage[cmex10]{amsmath}
\usepackage{amsfonts}
\usepackage{amssymb}
\usepackage{ntheorem}
\usepackage{dsfont}    %mathds offers better set-symbols for N,Z,Q,R,C
\usepackage{mathtools} %\coloneqq for perfect :=
\usepackage{stackengine}

\definecolor{PU_orange}{HTML}{EE7F2D}
\definecolor{PU_darkorange}{HTML}{994400}
\definecolor{PU_lightorange}{HTML}{FFAA66}
\definecolor{PU_black}{HTML}{000000}
\definecolor{PU_darkgray}{HTML}{7F7F83}
\definecolor{PU_lightgray}{HTML}{BDBEC1}

\DeclarePairedDelimiter\parentheses{\lparen}{\rparen}
\usepackage{algorithmic}
\usepackage{array}
\usepackage[tight,footnotesize]{subfigure}

\theoremseparator{.}

\newtheorem{prop}{Proposition}
\newtheorem{lem}{Lemma}
\newtheorem{theorem}{Theorem}
\theorembodyfont{\upshape}
\theoremseparator{.}

\newtheorem{remark}{Remark}
\newtheorem*{example*}{Example}

\newcommand{\eu}{\mathrm{e}}

\newcommand{\E}{\mathbb{E}}

\newcommand{\X}{\mathbf{X}}
\newcommand{\Z}{\mathbf{Z}}
\newcommand{\Y}{\mathbf{Y}}

\newcommand{\U}{\mathbf{U}}

\newcommand{\supp}{{\mathsf{supp}}}

\newcommand{\pam}{{\mathsf{PAM}}}

\newcommand\barbelow[1]{\stackunder[1.2pt]{$#1$}{\rule{0.9ex}{.075ex}}}

\long\def\symbolfootnote[#1]#2{\begingroup%
	\def\thefootnote{\fnsymbol{footnote}}\footnote[#1]{#2}\endgroup} %unnumbered footnote for thanks
\title{Upper and Lower Bounds on the Capacity of Amplitude-Constrained MIMO Channels\vspace{10pt}}% 
\author{\IEEEauthorblockN{Alex~Dytso\IEEEauthorrefmark{1}, Mario~Goldenbaum\IEEEauthorrefmark{1}, Shlomo Shamai (Shitz)\IEEEauthorrefmark{2},\\ and H. Vincent~Poor\IEEEauthorrefmark{1}\\ \vspace{7pt}}%
{\normalsize\IEEEauthorblockA{\IEEEauthorrefmark{1}Department of Electrical Engineering, Princeton University\\}%
\IEEEauthorblockA{\IEEEauthorrefmark{2}Department of Electrical Engineering, Technion -- Israel Institute of Technology}}}%%
\begin{document}
\maketitle
\symbolfootnote[0]{This work was supported in part by the U. S. National Science Foundation under Grants CCF-1420575 and CNS-1456793, by the German Research Foundation under Grant GO 2669/1-1, and by the 
European Union's Horizon 2020 Research And Innovation Programme, grant agreement no. 694630.}
%
%%%%%%%%%%%%%%%%%%%%%%%%%%%%%%%%%%%%%%%%%%%%%%%%%%%%%%%%%%%%%%%%%%%%%%%%%%%%%%%%%%%%%%%%%%%%%%%%%%%%%%%%%%%%%%%%%%%%%%%%%%%
%%%%%%%%%%%%%%%%%%%%%%%%%%%%%%%%%%%%%%%%%%%%%%%%%%%%%%%%%%%%%%%%%%%%%%%%%%%%%%%%%%%%%%%%%%%%%%%%%%%%%%%%%%%%%%%%%%%%%%%%%%%
%	Abstract
%%%%%%%%%%%%%%%%%%%%%%%%%%%%%%%%%%%%%%%%%%%%%%%%%%%%%%%%%%%%%%%%%%%%%%%%%%%%%%%%%%%%%%%%%%%%%%%%%%%%%%%%%%%%%%%%%%%%%%%%%%%
%%%%%%%%%%%%%%%%%%%%%%%%%%%%%%%%%%%%%%%%%%%%%%%%%%%%%%%%%%%%%%%%%%%%%%%%%%%%%%%%%%%%%%%%%%%%%%%%%%%%%%%%%%%%%%%%%%%%%%%%%%%
\begin{abstract}%
	In this work, novel upper and lower bounds for the capacity of channels with arbitrary constraints on the support of the channel input symbols are derived. As an immediate practical application, the case of multiple-input multiple-output channels with \emph{amplitude constraints} is considered. The bounds are shown to be within a constant gap if the channel matrix is invertible and are tight in the high amplitude regime for arbitrary channel matrices. Moreover, in the high amplitude regime, it is shown that the capacity scales linearly with the minimum between the number of transmit and receive antennas, similarly to the case of average power-constrained inputs.  
\end{abstract}%
%
%
%
%%%%%%%%%%%%%%%%%%%%%%%%%%%%%%%%%%%%%%%%%%%%%%%%%%%%%%%%%%%%%%%%%%%%%%%%%%%%%%%%%%%%%%%%%%%%%%%%%%%%%%%%%%%%%%%%%%%%%%%%%%%
%%%%%%%%%%%%%%%%%%%%%%%%%%%%%%%%%%%%%%%%%%%%%%%%%%%%%%%%%%%%%%%%%%%%%%%%%%%%%%%%%%%%%%%%%%%%%%%%%%%%%%%%%%%%%%%%%%%%%%%%%%%
%	Introduction
%%%%%%%%%%%%%%%%%%%%%%%%%%%%%%%%%%%%%%%%%%%%%%%%%%%%%%%%%%%%%%%%%%%%%%%%%%%%%%%%%%%%%%%%%%%%%%%%%%%%%%%%%%%%%%%%%%%%%%%%%%%
%%%%%%%%%%%%%%%%%%%%%%%%%%%%%%%%%%%%%%%%%%%%%%%%%%%%%%%%%%%%%%%%%%%%%%%%%%%%%%%%%%%%%%%%%%%%%%%%%%%%%%%%%%%%%%%%%%%%%%%%%%%
\section{Introduction}\label{sec:intro}%
While the capacity of a multiple-input multiple-output (MIMO) channel with an average power constraint is well understood \cite{Telatar:1999}, surprisingly, little is known about the capacity of the more practically relevant case in which the channel inputs are subject to \emph{amplitude constraints}.
%
%
%
%%%%%%%%%%%%%%%%%%%%%%%%%%%%%%%%%%%%%%%%%%%%%%%%%%%%%%%%%%%%%%%%%%%%%%%%%%%%%%%%%%%%%%%%%%%%%%%%%%%%%%%%%%%%%%%%%%%%%%%%%%%%
%\subsection{Related Work}
%
The first major contribution to this problem was a seminal work of Smith \cite{smith1971information}, in which it was shown that, for the scalar Gaussian noise channel with an amplitude-constraint, the capacity achieving input is discrete with finite support. 
%Smith in a seminal work in \cite{smith1971information} showed that for the scalar Gaussian noise channel the capacity achieving inputs are discrete with finite support. 
In \cite{ShamQuadrat}, this result was extended to peak-power-limited quadrature Gaussian channels. Using the approach of \cite{ShamQuadrat}, in \cite{rassouli2016capacity} the optimal input distribution was shown to be discrete for MIMO channels with an identity channel matrix and a Euclidian norm constraint on the input vector. Even though the optimal input distribution is known to be discrete, very little is known about the number or the optimal positions of the corresponding constellation points. To the best of our knowledge, the only exception is the work of \cite{sharma2010transition} in which for a scalar Gaussian noise channel it was shown that two point masses are optimal for amplitude values smaller than $1.671$ and three for amplitude values of up to $2.786$.

Using a dual capacity expression, in \cite{mckellips2004simple} McKellips derived an upper bound on the capacity of a scalar amplitude-constrained channel that is asymptotically tight in the high amplitude regime. By using a clever choice of an auxiliary channel output distribution in the dual capacity expression, the authors of \cite{thangaraj2015capacity} sharpened McKellips' bound and extended it to parallel MIMO channels with a Euclidian norm constraint on the input. The scalar version of the upper bound in \cite{thangaraj2015capacity} has been further sharpened in \cite{rassouli2016upper} by yet another choice of auxiliary output distribution.
In \cite{elmoslimany2016capacity}, asymptotic lower and upper bounds for a $2 \times 2$ MIMO system were presented and the gap between the bounds was specified. 

In this work, we make progress on this open problem by deriving several new upper and lower bounds that hold for channels with \emph{arbitrary constraints} on the support of the input distribution. We then apply them to the special case of MIMO channels with amplitude-constrained inputs. 
%
%
%
%%%%%%%%%%%%%%%%%%%%%%%%%%%%%%%%%%%%%%%%%%%%%%%%%%%%%%%%%%%%%%%%%%%%%%%%%%%%%%%%%%%%%%%%%%%%%%%%%%%%%%%%%%%%%%%%%%%%%%%%%%%
\subsection{Contributions and Paper Outline}
Our contributions and paper outline are as follows. The problem is stated in Section~\ref{sec:problem}. In Section~\ref{sec:UpperAndLowerBounds}, we derive upper and lower bounds on the capacity of a MIMO channel with an arbitrary constraint on the support of the input. In Section~\ref{sec:inv_channel}, we evaluate the performance of our bounds by studying MIMO channels with invertible channel matrices. In particular, in Theorem~\ref{thm:PackingEfficiecy} it is shown that our upper and lower bounds are within $n\log(\rho)$ bits, where $\rho$ is the packing efficiency and $n$ is the number of antennas. For diagonal channel matrices, it is shown in Theorem~\ref{thm:PAMparallel} that the Cartesian product of pulse-amplitude modulation (PAM) constellations achieves the capacity to within $1.64n$ bits. Section~\ref{sec:SingularValuesDecom} is devoted to MIMO channels with arbitrary channel matrices. It is shown that in the high amplitude regime, similarly to the average power-constrained channel, the capacity scales linearly with the minimum of the number of transmit and receive antennas. Section~\ref{sec:conclusion} concludes the paper.
%
%
%
%%%%%%%%%%%%%%%%%%%%%%%%%%%%%%%%%%%%%%%%%%%%%%%%%%%%%%%%%%%%%%%%%%%%%%%%%%%%%%%%%%%%%%%%%%%%%%%%%%%%%%%%%%%%%%%%%%%%%%%%%%%
\subsection{Notation}%
Vectors are denoted by bold lowercase letters, random vectors by bold uppercase letters, and matrices by bold uppercase sans serif letters (e.g., $\mathbf{x}$, $\mathbf{X}$, $\boldsymbol{\mathsf{X}}$). For any deterministic vector $\mathbf{x}\in\mathbb{R}^n$, $n\in\mathbb{N}$, we denote the Euclidian norm of $\mathbf{x}$ by $\|\mathbf{x}\|$. For some $\X\in\supp(\X)\subseteq\mathbb{R}^n$ and any $p>0$ we define
\begin{equation}%
	\|\X\|_p^p\coloneqq\frac{1}{n}\E[\|\X\|^p]\;,
	\label{eq:NormDef}%
\end{equation}%
where $\supp(\X)$ denotes the support of $\X$. Note that for $p \ge 1$, the quantity in \eqref{eq:NormDef} defines a norm. The norm of a matrix $\boldsymbol{\mathsf{H}}\in\mathbb{R}^{n\times n}$ is defined as
\begin{equation*}%
 \|\boldsymbol{\mathsf{H}}\|\coloneqq\sup_{\mathbf{x}:\mathbf{x}\neq\mathbf{0}}\frac{\|\boldsymbol{\mathsf{H}}\mathbf{x}\|}{\|\mathbf{x}\|}\;. 
\end{equation*}%

Let $\mathcal{S}$ be a subset of $\mathbb{R}^n$. Then,  
\begin{equation*}%
	\mathrm{Vol}(\mathcal{S})\coloneqq\int_{\mathcal{S}}\,\mathrm{d}\mathbf{x}
\end{equation*}%
denotes its volume.

Let $\mathbb{R}_+\coloneqq\{x\in\mathbb{R}:x\geq 0\}$. We define an $n$-dimensional ball or radius $r\in\mathbb{R}_+$ centered at $\mathbf{x}\in\mathbb{R}^n$ as the set
\begin{equation*}%
	\mathcal{B}_{\mathbf{x}}(r)\coloneqq\{\mathbf{y}:\|\mathbf{x}-\mathbf{y}\|\le r\}\;. 
\end{equation*}%
Recall that for any $\mathbf{x}\in\mathbb{R}^n$ and $r\in\mathbb{R}_+$,
\begin{equation*}%
	\mathrm{Vol}\bigl(\mathcal{B}_{\mathbf{x}}(r)\bigr)=\frac{\pi^{\frac{n}{2}}}{\Gamma\parentheses[\big]{\frac{n}{2}+1}}r^n\;.
\end{equation*}%

For any matrix $\boldsymbol{\mathsf{H}}\in\mathbb{R}^{k\times n}$ and some $\mathcal{S}\subset\mathbb{R}^n$ we define
\begin{equation*}%
	\boldsymbol{\mathsf{H}}\mathcal{S}\coloneqq\{\mathbf{y}:\mathbf{y}=\boldsymbol{\mathsf{H}}\mathbf{x}\,,\,\mathbf{x}\in\mathcal{S}\}\;. 
\end{equation*}%
Note that for an invertible $\boldsymbol{\mathsf{H}}\in\mathbb{R}^{n\times n}$ we have  
\begin{equation*}%
	\mathrm{Vol}(\boldsymbol{\mathsf{H}}\mathcal{S})=|\!\det(\boldsymbol{\mathsf{H}})|\mathrm{Vol}(\mathcal{S})\;. 
\end{equation*}% 
We define the maximum and minimum radius of a set $\mathcal{S}\subset\mathbb{R}^n$ that contains the origin as
\begin{align*}%
	r_{\mathsf{max}}(\mathcal{S})&\coloneqq\min\{r\in\mathbb{R}_+:\mathcal{S}\subset\mathcal{B}_{\mathbf{0}}(r)\}\;,\\
	%\radmax(\mathsf{S}) &= \frac{1}{2} \max_{{\bf x}_j , {\bf x}_i \in\mathsf{S}} \|  {\bf x}_j - {\bf x}_i \|  =  \min \left\{ r :  \mathcal{X} \subset \mathcal{B}_{ \bf x}(r)   \right\},\\
	%\radmin(\mathsf{S})&= \max \left\{ r : \mathcal{B}_{ \bf x}(r) \subseteq \mathcal{X}  \right\}.\\
	r_{\mathsf{min}}(\mathcal{S})&\coloneqq\max\{r\in\mathbb{R}_+:\mathcal{B}_{\mathbf{0}}(r)\subseteq\mathcal{S}\}\;.
\end{align*}%

For a given vector $\mathbf{a}=(a_1,\dots,a_n)\in\mathbb{R}^n_+$ we define
\begin{equation*}%
	\mathrm{Box}(\mathbf{a})\coloneqq\{\mathbf{x}\in\mathbb{R}^n:|x_i|\le a_i, i=1,\dots,n\}
\end{equation*}%
and the smallest box containing a given set $\mathcal{S}\subset\mathbb{R}^n$ as
\begin{equation*}
	%\mathrm{Box}(\mathcal{S})\coloneqq\inf\{\mathrm{Box}(\mathbf{a}):\mathcal{S}\subseteq\mathrm{Box}(\mathbf{a}),\mathbf{a}\in\mathbb{R}_+^n\}\;.
	\mathrm{Box}(\mathcal{S})\coloneqq\inf\{\mathrm{Box}(\mathbf{a}):\mathcal{S}\subseteq\mathrm{Box}(\mathbf{a})\}\;,
\end{equation*}%
respectively. Finally, all logarithms are taken to the base $2$, $\log^+\parentheses*{x}\coloneqq\max\{\log(x),0\}$, $Q(x)$, $x\in\mathbb{R}$, denotes the Q-function, and $\delta_{\mathbf{x}}(\mathbf{y})$ the Kronecker delta, which is one for $\mathbf{x}=\mathbf{y}$ and zero otherwise. 
%
%
%
%%%%%%%%%%%%%%%%%%%%%%%%%%%%%%%%%%%%%%%%%%%%%%%%%%%%%%%%%%%%%%%%%%%%%%%%%%%%%%%%%%%%%%%%%%%%%%%%%%%%%%%%%%%%%%%%%%%%%%%%%%%
%%%%%%%%%%%%%%%%%%%%%%%%%%%%%%%%%%%%%%%%%%%%%%%%%%%%%%%%%%%%%%%%%%%%%%%%%%%%%%%%%%%%%%%%%%%%%%%%%%%%%%%%%%%%%%%%%%%%%%%%%%%
%	Problem Statement
%%%%%%%%%%%%%%%%%%%%%%%%%%%%%%%%%%%%%%%%%%%%%%%%%%%%%%%%%%%%%%%%%%%%%%%%%%%%%%%%%%%%%%%%%%%%%%%%%%%%%%%%%%%%%%%%%%%%%%%%%%%
%%%%%%%%%%%%%%%%%%%%%%%%%%%%%%%%%%%%%%%%%%%%%%%%%%%%%%%%%%%%%%%%%%%%%%%%%%%%%%%%%%%%%%%%%%%%%%%%%%%%%%%%%%%%%%%%%%%%%%%%%%%
\section{Problem Statement}\label{sec:problem}%
Consider a MIMO system with $n_t\in\mathbb{N}$ transmit and $n_r\in\mathbb{N}$ receive antennas. The corresponding $n_r$-dimensional channel output for a single channel use is of the form %$\mathbf{Y}=\boldsymbol{\mathsf{H}}\mathbf{X}+\mathbf{Z}$
\begin{equation*}%
	\mathbf{Y}=\boldsymbol{\mathsf{H}}\mathbf{X}+\mathbf{Z}\;,
\end{equation*}%
for some fixed channel matrix $\boldsymbol{\mathsf{H}}\in\mathbb{R}^{n_r\times n_t}$.\footnote{Considering a real-valued channel model is without loss of generality.} Here and hereafter, we assume $\Z\sim\mathcal{N}(\mathbf{0},\boldsymbol{\mathsf{I}}_{n_r})$ is independent of the channel input $\X\in\mathbb{R}^{n_t}$ and $\boldsymbol{\mathsf{H}}$ is known to both the transmitter and the receiver, where $\boldsymbol{\mathsf{I}}_{n_r}$ denotes the $n_r\times n_r$ identity matrix.

Now, in all that follows let $\mathcal{X}\subset\mathbb{R}^{n_t}$ be a \emph{convex and compact} channel input space that contains the origin (i.e., the length-$n_t$ zero vector) and let $F_{\mathbf{X}}$ denote the cumulative distribution function of $\mathbf{X}$. As of the writing of this paper, the capacity  
\begin{equation}%
	C(\mathcal{X},\boldsymbol{\mathsf{H}})\coloneqq\max_{F_{\X}:\X\in\mathcal{X}}I(\X;\boldsymbol{\mathsf{H}}\X +\Z)
	\label{eq:capacity_amp_constr} 
\end{equation}%
of this channel is unknown and we are interested in finding novel lower and upper bounds. Even though most of the results in this paper hold for arbitrary $\mathcal{X}$, we are mainly interested in the two most important special cases:
\begin{itemize}
	\item[(i)] per-antenna amplitude constraints; that is, $\mathcal{X}=\mathrm{Box}(\mathbf{a})$ for some given $\mathbf{a}=(A_1,\dots,A_{n_t})\in\mathbb{R}_+^{n_t}$,
	\item[(ii)] $n_t$-dimensional amplitude constraint; that is, $\mathcal{X}=\mathcal{B}_{\mathbf{0}}(A)$ for some given $A\in\mathbb{R}_+$.
\end{itemize}%
\begin{remark}%
	Note that determining the capacity of a MIMO channel with average per-antenna power constraints is also still an open problem and has been solved for some special cases only \cite{vu2011miso,tuninetti2014capacity,loyka2017capacity,b}.
\end{remark}%
%
%
%
%%%%%%%%%%%%%%%%%%%%%%%%%%%%%%%%%%%%%%%%%%%%%%%%%%%%%%%%%%%%%%%%%%%%%%%%%%%%%%%%%%%%%%%%%%%%%%%%%%%%%%%%%%%%%%%%%%%%%%%%%%%
%%%%%%%%%%%%%%%%%%%%%%%%%%%%%%%%%%%%%%%%%%%%%%%%%%%%%%%%%%%%%%%%%%%%%%%%%%%%%%%%%%%%%%%%%%%%%%%%%%%%%%%%%%%%%%%%%%%%%%%%%%%
%	Upper and Lower Bounds
%%%%%%%%%%%%%%%%%%%%%%%%%%%%%%%%%%%%%%%%%%%%%%%%%%%%%%%%%%%%%%%%%%%%%%%%%%%%%%%%%%%%%%%%%%%%%%%%%%%%%%%%%%%%%%%%%%%%%%%%%%%
%%%%%%%%%%%%%%%%%%%%%%%%%%%%%%%%%%%%%%%%%%%%%%%%%%%%%%%%%%%%%%%%%%%%%%%%%%%%%%%%%%%%%%%%%%%%%%%%%%%%%%%%%%%%%%%%%%%%%%%%%%%
\section{Upper and Lower Bounds on the Capacity}\label{sec:UpperAndLowerBounds}%
%
%
%
%%%%%%%%%%%%%%%%%%%%%%%%%%%%%%%%%%%%%%%%%%%%%%%%%%%%%%%%%%%%%%%%%%%%%%%%%%%%%%%%%%%%%%%%%%%%%%%%%%%%%%%%%%%%%%%%%%%%%%%%%%%
\subsection{Upper Bounds}%
To establish our first upper bound on \eqref{eq:capacity_amp_constr}, we need the following result \cite[Th.\,1]{dytsoISIT2017ozarow}:
\begin{lem}\emph{(Maximum Entropy Under $p$-th Moment Constraint)}\label{thm:EntropyMax}%
	\;Let $n\in\mathbb{N}$ and $p\in(0,\infty)$ be arbitrary. Then, for any $\U\in\mathbb{R}^n$ such that $h(\U)<\infty$ and $\|\U\|_p<\infty$, we have
	\begin{equation*}%
		h(\U)\le n\log\parentheses*{k_{n,p}\,n^{\frac{1}{p}}\,\|\U\|_p}\;,
	\end{equation*}% 
	where
	\begin{equation*}%
	k_{n,p}\coloneqq\frac{\sqrt{\pi}\,\eu^{\frac{1}{p}}\parentheses[\big]{\frac{p}{n}}^{\frac{1}{p}}\Gamma\parentheses[\big]{\frac{n}{p}+1}^{\frac{1}{n}}}{\Gamma\parentheses[\big]{\frac{n}{2}+1}^{\frac{1}{n}}}\;.
	\label{eq: constant for moment entropy inequality}%
	\end{equation*}%
\end{lem}%
\begin{theorem}\emph{(Moment Upper Bound)}\label{thm:UpperBoundMaxEntropy}%
	\;For any channel input space $\mathcal{X}$ and any fixed channel matrix $\boldsymbol{\mathsf{H}}$, we have
	\begin{align*}%
		C(\mathcal{X},\boldsymbol{\mathsf{H}})\le\bar{C}_{\mathsf{M}}(\mathcal{X},\boldsymbol{\mathsf{H}})\coloneqq\inf_{p>0}n_r\log\parentheses*{\!\frac{k_{n_r,p} }{(2\pi\eu)^{\frac{1}{2}}}\,n_r^{\frac{1}{p}}\|\tilde{\mathbf{x}}+\Z\|_p},
		%\label{eq:UpperBound}
	\end{align*}%
	where $\tilde{\mathbf{x}}\in\boldsymbol{\mathsf{H}}\mathcal{X}$ is chosen such that $\|\tilde{\mathbf{x}}\|=r_{\mathsf{max}}(\boldsymbol{\mathsf{H}}\mathcal{X})$.
\end{theorem}%
\begin{IEEEproof}%
	Expressing \eqref{eq:capacity_amp_constr} in terms of differential entropies results in
	\begin{align}%
		%C(\mathcal{X},\boldsymbol{\mathsf{H}})&\stackrel{a)}{=}\max_{F_{\X}:\X\in\mathcal{X},\|\X\|_p\le r_{\mathsf{max}}(\mathcal{X})}I(\X;\Y)\nonumber\\
		\hspace{-5pt}C(\mathcal{X},\boldsymbol{\mathsf{H}})&=\max_{F_{\X}:\X\in\mathcal{X}}h(\boldsymbol{\mathsf{H}}\X+\Z)-h(\Z)\nonumber\\
		%&\max_{ \X:   \X \in\mathcal{X}} I(\X; \boldsymbol{\mathsf{H}}\, \X +\Z) \\
		%& \stackrel{a)}{=}  \max_{ \X:   \X \in\mathcal{X},  \| \X \|_p \le {\rm rad} (\mathsf{S})} I(\X; \boldsymbol{\mathsf{H}}\, \X +\Z) \\
		%&\stackrel{a)}{\le}\max_{F_{\X}:\X\in\mathcal{X}}n_r\log\parentheses*{k_{n_r,p}\cdot n_r^{\frac{1}{p}}\cdot\|\boldsymbol{\mathsf{H}}\X+\Z\|_p}\nonumber\\
		%&\hspace{12pt}-\frac{n_r}{2}\log(2\pi\eu)\nonumber\\
		&\stackrel{a)}{\le}\max_{F_{\X}:\X\in\mathcal{X}}n_r\log\parentheses*{\frac{k_{n_r,p}}{(2\pi\eu)^{\frac{1}{2}}}\,n_r^{\frac{1}{p}}\,\|\boldsymbol{\mathsf{H}}\X+\Z\|_p}\nonumber\\
		& \stackrel{b)}{=}n_r\log\parentheses*{\frac{k_{n_r,p} }{(2\pi\eu)^{\frac{1}{2}}}\,n_r^{\frac{1}{p}}\max_{F_{\X}:\X\in\mathcal{X}}\|\boldsymbol{\mathsf{H}}\X+\Z\|_p}\!,\label{eq:moment_bound1}
		%& \stackrel{d)}{=} \log \left( \frac{k_{n,p} }{ (2 \pi \eu)^{\frac{p}{2}}}  \cdot n^{\frac{1}{p}} \cdot    \max_{    {\bf x} \in\mathcal{X}}   \| \boldsymbol{\mathsf{H}} {\bf x}+\Z\|_p \right)
	\end{align}%
	where $a)$ follows from Lemma~\ref{thm:EntropyMax} with the fact that $h(\Z)=\frac{n_r}{2}\log(2\pi\eu)$ and $b)$ from the monotonicity of the logarithm.

	Now, notice that $\|\boldsymbol{\mathsf{H}}\X+\Z\|_p$ is linear and bounded (and therefore continuous) in $F_{\X}$ so that it attains its maximum at an extreme point of the set $\mathcal{F}_{\X}\coloneqq\{F_{\X}:\X\in\mathcal{X}\}$ (i.e., the set of all cumulative distribution functions of $\X$). As a matter of fact \cite{witsenhausen1980convexity}, the extreme points of $\mathcal{F}_{\X}$ are given by the set of degenerate distributions on $\mathcal{X}$; that is, $\{F_{\X}(\mathbf{y})=\delta_{\mathbf{x}}(\mathbf{y}),\mathbf{y}\in\mathcal{X}\}_{\mathbf{x}\in\mathcal{X}}$. This allows us to conclude
	\begin{equation*}%
		\max_{F_{\X}:\X\in\mathcal{X}}\|\boldsymbol{\mathsf{H}}\X+\Z\|_p=\max_{\mathbf{x}\in\mathcal{X}}\|\boldsymbol{\mathsf{H}}\mathbf{x}+\Z\|_p\;. 
		\label{eq:Bound2}
	\end{equation*}% 
	Observe that the Euclidian norm is a convex function, which is therefore maximized at the boundary of the set $\boldsymbol{\mathsf{H}}\mathcal{X}$. Combining this with \eqref{eq:moment_bound1} and taking the infimum over $p>0$ completes the proof.
\end{IEEEproof}%
%
%\begin{remark}
%Observe, that the idea of   the bound in Theorem~\ref{thm:UpperBoundMaxEntropy} is  to add  a redundant constraint in the form of $p$-th moment.  An interesting future direction is to generalization this upper bound by %selecting the best redundant constraint. In other words, solving the following optimization problem
%\begin{align}
%\inf_{f} \max_{X:  \E[ f(X) ]  \le \max_{ y: |y| \le A}  f(y)} I(X; X+Z).
%\end{align}
%\end{remark}

The following theorem provides two alternative upper bounds that are based on duality arguments.
\begin{theorem}\emph{(Duality Upper Bounds)}\label{thm:DualityBounds} 
	\;For any channel input space $\mathcal{X}$ and any fixed channel matrix $\boldsymbol{\mathsf{H}}$
		\begin{equation}%
			C(\mathcal{X},\boldsymbol{\mathsf{H}})\le\bar{C}_{\mathsf{Dual},1}(\mathcal{X},\boldsymbol{\mathsf{H}})\coloneqq\log\parentheses*{c_{n_r}(d) + \frac{\mathrm{Vol}\parentheses[\big]{\mathcal{B}_{\mathbf{0}}(d)}}{(2\pi\eu)^{\frac{n_r}{2}}}}, 
			\label{eq:VolUpperBound}
		\end{equation}%
		where
		\begin{equation*}%
			d\coloneqq r_{\mathsf{max}}(\boldsymbol{\mathsf{H}}\mathcal{X})\,,\,c_{n_r}(d)\coloneqq\sum_{i=1}^{n_r-1} {{n_r-1}\choose{i}}\frac{\Gamma\parentheses*{\frac{n_r-1}{2}}}{2^{\frac{n_r}{2}}\Gamma\parentheses*{ \frac{n_r}{2}}} d^i\;,
		\end{equation*}%
		and
		\begin{equation}%
			C(\mathcal{X},\boldsymbol{\mathsf{H}})\le\bar{C}_{\mathsf{Dual},2} (\mathcal{X},\boldsymbol{\mathsf{H}})\coloneqq\sum_{i=1}^{n_r}\log\parentheses*{1+\frac{2A_i}{\sqrt{2\pi\eu}}}\;, 
			\label{eq:ProjecLikeBound}
		\end{equation}%
		where $\mathbf{a}=(A_1,\dots,A_{n_r})$ such that $\mathrm{Box}(\mathbf{a})=\mathrm{Box}(\boldsymbol{\mathsf{H}}\mathcal{X})$.
\end{theorem}%
\begin{IEEEproof}
	 Using duality bounds, it has been shown in \cite{thangaraj2015capacity} that for any centered $n$-dimensional ball of radius $r\in\mathbb{R}_+$% (i.e., $\mathcal{B}_{\mathbf{0}}(r)$) it holds 
	\begin{equation}%
		\max_{F_{\X}:\X\in\mathcal{B}_{\mathbf{0}}(r)}I(\X;\X+\Z)\le\log\parentheses*{c_n(r)+\frac{\mathrm{Vol}\parentheses[\big]{\mathcal{B}_{\mathbf{0}}(r)}}{(2\pi\eu)^{\frac{n}{2}}}}\;,
		\label{eq:KramerUpperBound}
	\end{equation}% 
	where $c_n(r)\coloneqq\sum_{i=1}^{n-1}{{n-1}\choose{i}}\frac{\Gamma\parentheses*{\frac{n-1}{2}}}{2^{\frac{n}{2}}\Gamma\parentheses*{\frac{n}{2}}}r^i$. 

	Now, observe that
	\begin{align}%
		%\max_{\X\in\mathcal{X}}I(\X;\boldsymbol{\mathsf{H}}\X+\Z) &=\max_{\X\in\mathcal{X}}h(\boldsymbol{\mathsf{H}}\X+\Z)-h(\Z)\notag\\
		C(\mathcal{X},\boldsymbol{\mathsf{H}})&=\max_{F_{\X}:\X\in\mathcal{X}}h(\boldsymbol{\mathsf{H}}\X+\Z)-h(\boldsymbol{\mathsf{H}}\X+\Z|\boldsymbol{\mathsf{H}}\X)\notag\\
		%&=\max_{\X\in\mathcal{X}}h(\boldsymbol{\mathsf{H}}\X+\Z) - h(\boldsymbol{\mathsf{H}} \X+\Z| \boldsymbol{\mathsf{H}} \X ) \notag\\
		&=\max_{F_{\X}:\X\in\mathcal{X}}I(\boldsymbol{\mathsf{H}}\X;\boldsymbol{\mathsf{H}}\X+\Z) \notag\\
		&=\max_{F_{\tilde{\X}}:\tilde{\X}\in\boldsymbol{\mathsf{H}}\mathcal{X}}I(\tilde{\X};\tilde{\X}+\Z) \label{eq:Dioganzlization}\\
		&\stackrel{a)}{\le}\max_{F_{\tilde{\X}}:\tilde{\X}\in\mathcal{B}_{\mathbf{0}}(d),d\coloneqq r_{\mathsf{max}}(\boldsymbol{\mathsf{H}}\mathcal{X})}I(\tilde{\X};\tilde{\X}+\Z)\notag\\
		&\stackrel{b)}{\le}\log\parentheses*{c_{n_r}(d)+\frac{\mathrm{Vol}\parentheses[\big]{\mathcal{B}_{\mathbf{0}}(d)}}{(2\pi\eu)^{\frac{n_r}{2}}}}\notag\;.
	\end{align}%
	Here, $a)$ follows from enlarging the optimization domain and $b)$ from using the upper bound in \eqref{eq:KramerUpperBound}. This proves \eqref{eq:VolUpperBound}. 
	
	In order to show the upper bound in \eqref{eq:ProjecLikeBound}, we proceed with an alternative upper bound to \eqref{eq:Dioganzlization}:
	\begin{align*}%
		C(\mathcal{X},\boldsymbol{\mathsf{H}})&=\max_{F_{\tilde{\X}}:\tilde{\X}\in\boldsymbol{\mathsf{H}}\mathcal{X}}I(\tilde{\X};\tilde{\X}+\Z)\\
		&\stackrel{a)}{\le}\max_{F_{\tilde{\X}}:\tilde{\X}\in\mathrm{Box}(\boldsymbol{\mathsf{H}}\mathcal{X})}I(\tilde{\X};\tilde{\X}+\Z)\\
		&\stackrel{b)}{\le}\max_{F_{\tilde{\X}}:\tilde{\X}\in\mathrm{Box}(\boldsymbol{\mathsf{H}}\mathcal{X})}\sum_{i=1}^{n_r}I(\tilde{X}_i;\tilde{X}_i+Z_i)\\
		&\stackrel{c)}{=}\sum_{i=1}^{n_r}\max_{F_{\tilde{X}_i}:|\tilde{X}_i|\le A_i}I(\tilde{X}_i;\tilde{X}_i+Z_i)\\
		&\stackrel{d)}{\le}\sum_{i=1}^{n_r}\log\parentheses*{1 +\frac{2A_i}{\sqrt{2\pi\eu}}}\;,
	\end{align*}%
	where the (in)equalities follow from: $a)$ enlarging the optimization domain; $b)$ single-letterizing the mutual information; $c)$ choosing individual amplitude constraints $(A_1,\dots,A_{n_r})\eqqcolon\mathbf{a}\in\mathbb{R}_+^{n_r}$ such that $\mathrm{Box}(\mathbf{a})=\mathrm{Box}(\boldsymbol{\mathsf{H}}\mathcal{X})$; and $d)$ using the upper bound in \eqref{eq:KramerUpperBound} for $n=1$. This concludes the proof. 
\end{IEEEproof}%
%
%
%In Section~XXXXXX, we present a comparison of the upper bounds of Theorems~\ref{thm:UpperBoundMaxEntropy} and \ref{thm:DualityBounds} by means of simple example. 
%
%
%
%%%%%%%%%%%%%%%%%%%%%%%%%%%%%%%%%%%%%%%%%%%%%%%%%%%%%%%%%%%%%%%%%%%%%%%%%%%%%%%%%%%%%%%%%%%%%%%%%%%%%%%%%%%%%%%%%%%%%%%%%%%
\subsection{Lower Bounds} 
A classical approach to bound a mutual information from below is to use the entropy power inequality (EPI).  
\begin{theorem}\emph{(EPI Lower Bounds)}\label{thm:LowerBound} 
	\;For any fixed channel matrix $\boldsymbol{\mathsf{H}}$ and any channel input space $\mathcal{X}$ with $\X$ absolutely continuous, we have
	\begin{equation}%
		C(\mathcal{X},\boldsymbol{\mathsf{H}})\ge\barbelow{C}_{\mathsf{EPI}}(\mathcal{X},\boldsymbol{\mathsf{H}})\coloneqq\max_{F_{\X}:\X\in\mathcal{X}}\frac{n_r}{2}\log\parentheses*{\!1+\frac{2^{\frac{2}{n_r}h(\boldsymbol{\mathsf{H}}\X)}}{2\pi\eu}\!}. \label{eq:lowerbound:EPI}
	\end{equation}% 
	Moreover, if $n_t=n_r=n$, $\boldsymbol{\mathsf{H}}\in\mathbb{R}^{n\times n}$ is invertible, and $\X$ is uniformly distributed over $\mathcal{X}$, then	
	\begin{equation}%
		C(\mathcal{X},\boldsymbol{\mathsf{H}})\ge\barbelow{C}_{\mathsf{EPI}}(\mathcal{X},\boldsymbol{\mathsf{H}})\coloneqq\frac{n}{2}\log\parentheses*{\!1+\frac{|\mathrm{det}(\boldsymbol{\mathsf{H}})|^{\frac{2}{n}}\mathrm{Vol}(\mathcal{X})^{\frac{2}{n}}}{2\pi\eu}\!}.
		\label{eq:AssumingInvertability}
	\end{equation}% 
\end{theorem}%
\begin{IEEEproof}%
	By means of the EPI
	\begin{equation*}%
		2^{\frac{2}{n_r}h(\boldsymbol{\mathsf{H}}\X+\Z)}\ge 2^{\frac{2}{n_r}h(\boldsymbol{\mathsf{H}}\X)}+2^{\frac{2}{n_r}h(\Z)}\;,
		%&={\rm Vol}  \left( \boldsymbol{\mathsf{H}} \cdot \mathcal{X}\right)^{\frac{2}{n_r} } +2 \pi \eu.
	\end{equation*}%
	we conclude
	\begin{equation}
		2^{\frac{2}{n_r}C(\mathcal{X},\boldsymbol{\mathsf{H}})}\ge 1+(2\pi\eu)^{-1}2^{\frac{2}{n_r}\displaystyle{\max_{F_{\X}:\X\in\mathcal{X}}}h(\boldsymbol{\mathsf{H}}\X)}\;,
		\label{eq:epi1}%
	\end{equation}%
	which finalizes the proof of the lower bound in \eqref{eq:lowerbound:EPI}.

	To show the lower bound in \eqref{eq:AssumingInvertability}, all we need is to recall that
	\begin{equation*}%
		h(\boldsymbol{\mathsf{H}}\X)=h(\X)+\log|\mathrm{det}(\boldsymbol{\mathsf{H}})|\;,
	\end{equation*}%
	which is maximized for $\X$ uniformly distributed over $\mathcal{X}$. But if $\X$ is uniformly drawn from $\mathcal{X}$, we have
	%
%	 let $\X$ be uniformly distributed over $\mathcal{X}$. As $\boldsymbol{\mathsf{H}}$ is full rank, $\boldsymbol{\mathsf{H}}\X$ is uniformly distributed over $\boldsymbol{\mathsf{H}}\mathcal{X}$ with density
%	\begin{equation*}%
%		f_{\boldsymbol{\mathsf{H}}\X}(\mathbf{x})=\frac{1}{|\mathrm{det}(\boldsymbol{\mathsf{H}})|\mathrm{Vol}(\mathcal{X})}\;,\;\mathbf{x}\in\boldsymbol{\mathsf{H}}\mathcal{X}\;.
%	\end{equation*}% 
%	This, however, implies that
	\begin{equation*}%
		2^{\frac{2}{n}h(\boldsymbol{\mathsf{H}}\X)}=\mathrm{Vol}(\boldsymbol{\mathsf{H}}\mathcal{X})^{\frac{2}{n}}=|\mathrm{det}(\boldsymbol{\mathsf{H}})|^{\frac{2}{n}}\mathrm{Vol}(\mathcal{X})^{\frac{2}{n}}\;,
	\end{equation*}%
	which completes the proof.
	%Observe that $F_{\X}$ chosen to be uniform over $\mathcal{X}$ maximizes the right-hand side of \eqref{eq:epi1}, which in combination with the EPI results in the lower bound \eqref{eq:AssumingInvertability}.
	%\begin{align}
	%I(\X; \boldsymbol{\mathsf{H}}\, \X +\Z)  \ge  \frac{n}{2} \log  \left(1 +  \frac{{\rm Vol}  \left( \boldsymbol{\mathsf{H}} \cdot \mathcal{X}\right)^{\frac{2}{n} }}{2 \pi \eu}   \right).
	%\end{align} 
	%This concludes the proof. 
\end{IEEEproof}%

The results in \cite{smith1971information,ShamQuadrat,rassouli2016capacity} suggest that the channel input distribution that maximizes \eqref{eq:capacity_amp_constr} might be discrete. Therefore, there is a need for lower bounds that, unlike the bounds in Theorem~\ref{thm:LowerBound}, rely on discrete inputs.
\begin{remark}%
	We note that the problem of finding the optimal input distribution of a general MIMO channel with an amplitude constraint is still open. The technical difficulty relies on the fact that the identity theorem from complex analysis, a key tool in the method developed by Smith \cite{smith1971information} for the scalar case and later used by \cite{chan2005MIMObounded} for the MIMO channel, does not extend to $\mathbb{R}^{n}$ and $\mathbb{C}^{n}$. The interested reader is referred to \cite{sommerfeld2008boundedness} for a detailed discussion on this issue with examples of why the identity theorem fails in the MIMO setting. 
\end{remark}%
\begin{theorem}\emph{(Ozarow-Wyner Type Lower Bound)}\label{prop:OWimproved}
	\;Let $\X_D\in\supp(\X_D)\subset\mathbb{R}^{n_t}$ be a discrete random vector of finite entropy, $g:\mathbb{R}^{n_r}\to\mathbb{R}^{n_t}$ a measurable function, and $p>0$. Furthermore, let $\mathcal{K}_p$ be a set of continuous random vectors, independent of $\X_D$, such that for every $\U \in\mathcal{K}_p$ we have $ h(\U)<\infty$, $\|\U\|_{p}<\infty$, and 
	\begin{equation}
		\supp (\U+{\bf  x}_i)  \cap  \supp (\U+{\bf  x}_j) =\varnothing
		\label{eq: assumption on U}
	\end{equation}%
	for all $\mathbf{x}_i,\mathbf{x}_j\in\supp(\X_D)$, $i\neq j$. Then, 
	\begin{equation*}
		C(\mathcal{X},\boldsymbol{\mathsf{H}})\ge\barbelow{C}_{\mathsf{OW}}(\mathcal{X},\boldsymbol{\mathsf{H}})\coloneqq [H(\X_D)-\mathsf{gap}]^{+}\;,
		%[H(\X_D)-\gap^{\star}]^{+}\le I(\X_D;\Y) \le H(\X_D)\;,
		\label{eq: OWbound}
	\end{equation*}% 
	where 
	\begin{equation*}
		\mathsf{gap}\coloneqq\inf_{\substack{\U\in\mathcal{K}_p\\g\,\textup{measurable}\\p>0}}\parentheses[\big]{G_{1,p}(\U,\X_D,g) + G_{2,p}(\U)}
		\label{eq:gap}
	\end{equation*}%
	with
	\begin{align}
		G_{1,p}(\U,\X_D,g)&\coloneqq  n_t\log\parentheses*{\frac{\|\U+\X_D-g(\Y)\|_p}{\|\U\|_p}}\;,\label{eq:G_1p}\\
		%& \stackrel{ \text{ for  $p \ge 1$} }{\le}     \log \left( 1+\frac{ \mmpe^{\frac{1}{p}}(\X_D,\snr,p) }{\| \U \|_p}  \right), \\
		G_{2,p}(\U)&\coloneqq n_t\log\parentheses*{\frac{k_{n_t,p}\,n_t^{\frac{1}{p}}\,\|\U \|_p}{2^{\frac{1}{n_t}h(\U)}}}\;,\label{eq:G_2p}%
	\end{align}%
	and $k_{n_t,p}$ as defined in Lemma~\ref{thm:EntropyMax}, respectively.
\end{theorem}%
\begin{IEEEproof}%
	The proof is identical to \cite[Th.\,2]{dytsoISIT2017ozarow}. In order to make the manuscript more self-contained, we repeat it here.
	
	Let $\U$ and $\X_D$ be statistically independent. Then, the mutual information $I(\X_D;\Y)$ can be lower bounded~as
	\begin{align}
		I( \X_D; \Y) & \stackrel{a)}{\ge}I( \X_D+ \U;\Y)\nonumber \\
		&= h(\X_D+ \U)- h(\X_D+ \U| \Y) \nonumber\\
		& \stackrel{b)}{=} 
		H(\X_D) +h(\U)- h(\X_D+ \U| \Y)\;.
		\label{eq: OWproof-decomp}
	\end{align}%
	Here, $a)$ follows from the data processing inequality as $\X_D+\U\to\X_D\to \Y$ forms a Markov chain in that order and $b)$ from the assumption in \eqref{eq: assumption on U}. By using Lemma~\ref{thm:EntropyMax}, we have that the last term in \eqref{eq: OWproof-decomp} can be bounded from above as
	\begin{equation*}
		h(\X_D+\U|\Y)\le n_t\log\parentheses*{k_{n_t,p}\,n_t^{\frac{1}{p}}\,\|\X_D+\U-g(\Y)\|_p}\;.
		\label{eq: third term after triangular inequality}
	\end{equation*}%
	% 
	%where the last inequality follows by the triangle inequality which holds for $2p \ge 1$.
	Combining this expression with \eqref{eq: OWproof-decomp} results in 
	\begin{equation*}
		I(\X_D;\Y)\geq H(\X_D)-\bigl(G_{1,p}(\U,\X_D,g) + G_{2,p}(\U)\bigr)\;,
		\label{eq:OW_bound_gap}%
	\end{equation*}%
	with $G_{1,p}$ and $G_{2,p}$ as defined in \eqref{eq:G_1p} and \eqref{eq:G_2p}, respectively. Maximizing the right-hand side over all $\U\in\mathcal{K}_p$, measurable functions $g:\mathds{R}^{n_r}\to\mathds{R}^{n_t}$, and $p>0$ provides the bound.
\end{IEEEproof}%
Interestingly, the bound of Theorem~\ref{prop:OWimproved} holds for arbitrary channels and the interested reader is referred to \cite{dytsoISIT2017ozarow} for details. 

We conclude the section by providing a lower bound that is based on Jensen's inequality and holds for arbitrary inputs. 
\begin{theorem}\emph{(Jensen's Inequality Lower Bound)}\label{thm:Jensen'sBound} 
	\;For any channel input space $\mathcal{X}$ and fixed channel matrix $\boldsymbol{\mathsf{H}}$, we have
	\begin{equation*}%
		C(\mathcal{X},\boldsymbol{\mathsf{H}})\ge\barbelow{C}_{\mathsf{Jensen}}(\mathcal{X},\boldsymbol{\mathsf{H}})\coloneqq \max_{F_{\X}:\X\in\mathcal{X}}\log^+\parentheses*{ \left(\frac{2}{\eu} \right)^{\frac{n_r}{2}}\E\biggl[\exp\parentheses*{-\frac{\|\boldsymbol{\mathsf{H}}(\X-\X')\|^2}{4}}\biggl]^{-1}}\;,
		%\label{eq:JenIneqBound}%
	\end{equation*}%
	where $\X'$ is an independent copy of $\X$. 
	\end{theorem} 

\begin{IEEEproof}%
	In order to show the lower bound, we follow an approach of \cite{DytsoTINPublished}. Note that by Jensen's inequality
	\begin{equation}
		h(\Y)=-\E[\log f_{\Y}(\Y)]\ge -\log\E[f_{\Y}(\Y)]=-\log\int_{\mathbb{R}^{n_r}}\!f_{\Y}(\mathbf{y})f_{\Y}(\mathbf{y})\,\mathrm{d}\mathbf{y}\;.
		\label{eq:IneqJensens}%
	\end{equation}%
	Now, evaluating the integral in \eqref{eq:IneqJensens} results in
	\begin{align}%
		\int_{\mathbb{R}^{n_r}}f_{\Y}(\mathbf{y})f_{\Y}(\mathbf{y})\,\mathrm{d}\mathbf{y}&=\frac{1}{(2\pi)^{n_r}}\int_{\mathbb{R}^{n_r}}\E\biggl[\eu^{-\frac{\|\mathbf{y}-\boldsymbol{\mathsf{H}}\X\|^2}{2}}\biggr]\E\biggl[\eu^{-\frac{\|\mathbf{y}-\boldsymbol{\mathsf{H}}\X'\|^2}{2}}\biggr]\mathrm{d}\mathbf{y}\notag\\
		&\stackrel{a)}{=}\frac{1}{(2\pi)^{n_r}}\E\biggl[\int_{\mathbb{R}^{n_r}}\eu^{-\frac{\|\mathbf{y}-\boldsymbol{\mathsf{H}}\mathbf{X}\|^2+\|\mathbf{y}-\boldsymbol{\mathsf{H}}\mathbf{X}'\|^2}{2}}\mathrm{d}\mathbf{y}\biggr]\notag\\
		&\stackrel{b)}{=}\frac{1}{(2\pi)^{n_r}}\E\biggl[\eu^{-\frac{\|\boldsymbol{\mathsf{H}}\X-\boldsymbol{\mathsf{H}}\X'\|^2}{4}}\int_{\mathbb{R}^{n_r}}\eu^{-\|\mathbf{y}-\frac{\boldsymbol{\mathsf{H}} (\X-\X')}{2}\|^2 }\mathrm{d}\mathbf{y}\biggr] \notag\\
	    &\stackrel{c)}{=}\frac{1}{2^{n_r}\pi^\frac{n_r}{2}}\E\biggl[\eu^{-\frac{\|\boldsymbol{\mathsf{H}}(\X-\X')\|^2}{4}}\biggr]\;, \label{eq:FinalAfterJen} 
	 \end{align}% 
	where $a)$ follows from the independence of $\X$ and $\X'$ and Tonelli's theorem, $b)$ from completing a square, and $c)$ from the fact that $\int_{\mathbb{R}^{n_r}}\eu^{-\|\mathbf{y}-\frac{\boldsymbol{\mathsf{H}} (\X-\X')}{2}\|^2 }\mathrm{d}\mathbf{y}=\int_{\mathbb{R}^{n_r}}\eu^{-\|\mathbf{y}\|^2 }\mathrm{d}\mathbf{y}=\pi^{\frac{n_r}{2}}$.
 
	Finally, combining \eqref{eq:IneqJensens} and \eqref{eq:FinalAfterJen}, subtracting $h(\Z)=\frac{n_r}{2}\log(2\pi\eu)$, and maximizing over $F_{\X}$ proves the result. 
% \begin{align*}
% h(\Y) -h(\Z) &\ge \log \left( \frac{2^{\frac{n+1}{2}}  \pi^\frac{n}{2}}{  \E \left[    \eu^{-\frac{\|\boldsymbol{\mathsf{H}}  (\X-\X_1)\|^2}{4}}   \right]   }   \right) - \log ( 2 \pi \eu)^{\frac{n}{2}}\\
% &=  \log  \left( \frac{ \sqrt{2}  }{ \eu^{\frac{n}{2}}  \E \left[    \eu^{-\frac{\|\boldsymbol{\mathsf{H}}  (\X-\X_1)\|^2}{4}}   \right]   }   \right).
% \end{align*}
\end{IEEEproof}%
%
%
%
%%%%%%%%%%%%%%%%%%%%%%%%%%%%%%%%%%%%%%%%%%%%%%%%%%%%%%%%%%%%%%%%%%%%%%%%%%%%%%%%%%%%%%%%%%%%%%%%%%%%%%%%%%%%%%%%%%%%%%%%%%%
%%%%%%%%%%%%%%%%%%%%%%%%%%%%%%%%%%%%%%%%%%%%%%%%%%%%%%%%%%%%%%%%%%%%%%%%%%%%%%%%%%%%%%%%%%%%%%%%%%%%%%%%%%%%%%%%%%%%%%%%%%%
%	Invertible Channel Matrices
%%%%%%%%%%%%%%%%%%%%%%%%%%%%%%%%%%%%%%%%%%%%%%%%%%%%%%%%%%%%%%%%%%%%%%%%%%%%%%%%%%%%%%%%%%%%%%%%%%%%%%%%%%%%%%%%%%%%%%%%%%%
%%%%%%%%%%%%%%%%%%%%%%%%%%%%%%%%%%%%%%%%%%%%%%%%%%%%%%%%%%%%%%%%%%%%%%%%%%%%%%%%%%%%%%%%%%%%%%%%%%%%%%%%%%%%%%%%%%%%%%%%%%%
\section{Invertible Channel Matrices}\label{sec:inv_channel}%
Consider the case of $n_t=n_r=n$ antennas with $\boldsymbol{\mathsf{H}}\in\mathbb{R}^{n\times n}$ being invertible. In this section, we evaluate some of the lower and upper bounds given in the previous section for the special case of $\boldsymbol{\mathsf{H}}$ being also diagonal and then characterize the gap to the capacity for arbitrary invertible channel matrices. 
%
%
%
%%%%%%%%%%%%%%%%%%%%%%%%%%%%%%%%%%%%%%%%%%%%%%%%%%%%%%%%%%%%%%%%%%%%%%%%%%%%%%%%%%%%%%%%%%%%%%%%%%%%%%%%%%%%%%%%%%%%%%%%%%%
\subsection{Diagonal Channel Matrices}\label{sec:ParallelChannels}%
Suppose the channel inputs are subject to per-antenna or an $n$-dimensional amplitude constraint. Then, the duality upper bound $\bar{C}_{\mathsf{Dual},2}(\mathcal{X},\boldsymbol{\mathsf{H}})$ of Theorem~\ref{thm:DualityBounds} is of the following form.
\begin{theorem}\label{thm:UpperBoundsParallel}\emph{(Upper Bounds)}%
	\;Let $\boldsymbol{\mathsf{H}}=\mathrm{diag}(h_{11},\dots,h_{nn})\in\mathbb{R}^{n\times n}$ be fixed. If $\mathcal{X}=\mathrm{Box}(\mathbf{a})$ for some $\mathbf{a}=(A_1,\dots,A_n)\in\mathbb{R}_+^n$, then
	\begin{equation}%
		\bar{C}_{\mathsf{Dual},2}(\mathrm{Box}(\mathbf{a}),\boldsymbol{\mathsf{H}})=\sum_{i=1}^n\log\parentheses*{1+\frac{2 |h_{ii}|A_i}{\sqrt{2\pi\eu}}}\;.
		\label{eq:BoundDual2Box}
	\end{equation}%
	Moreover, if $\mathcal{X}=\mathcal{B}_{\mathbf{0}}(A)$ for some $A\in\mathbb{R}_+$, then
	\begin{equation}%
		\bar{C}_{\mathsf{Dual},2}(\mathcal{B}_{\mathbf{0}}(A),\boldsymbol{\mathsf{H}})=\sum_{i=1}^n\log\parentheses*{1+\frac{2|h_{ii}|A}{\sqrt{n}\sqrt{2\pi\eu}}}\;.
		\label{eq:BoundDual2BallDiagonal}
	\end{equation}%
\end{theorem}%
\begin{IEEEproof}%
	The bound in \eqref{eq:BoundDual2Box} immediately follows from Theorem~\ref{thm:DualityBounds} by observing that $\mathrm{Box}(\boldsymbol{\mathsf{H}}\mathrm{Box}(\mathbf{a}))=\mathrm{Box}(\boldsymbol{\mathsf{H}}\mathbf{a})$. The bound in \eqref{eq:BoundDual2BallDiagonal} follows from Theorem~\ref{thm:DualityBounds} by the fact that 
	\begin{equation*}
		\mathrm{Box}\parentheses[\big]{\boldsymbol{\mathsf{H}}\mathcal{B}_{\mathbf{0}}(A)}\subset\mathrm{Box}\parentheses*{\boldsymbol{\mathsf{H}}\mathrm{Box}\parentheses[\big]{\mathcal{B}_{\mathbf{0}}(A)}}=\mathrm{Box}(\mathbf{h})\;,
	\end{equation*}
	where $\mathbf{h}\coloneqq\frac{A}{\sqrt{n}}(|h_{11}|,\dots,|h_{nn}|)$. This concludes the proof. 
\end{IEEEproof}%

For an arbitrary channel input space $\mathcal{X}$, the EPI lower bound of Theorem~\ref{thm:LowerBound} and Jensen's inequality lower bound of Theorem~\ref{thm:Jensen'sBound} evaluate to the following.
\begin{theorem}\label{thm:lowerBoundsParallel}\emph{(Lower Bounds)}%
	\;Let $\boldsymbol{\mathsf{H}}=\mathrm{diag}(h_{11},\dots,h_{nn})\in\mathbb{R}^{n\times n}$ be fixed and $\mathcal{X}$ arbitrary. Then, 
	\begin{equation}
		\barbelow{C}_{\mathsf{Jensen}}(\mathcal{X},\boldsymbol{\mathsf{H}})=\log^+\parentheses*{\left(\frac{2}{\eu}\right)^{\frac{n}{2}}\frac{1}{\psi(\boldsymbol{\mathsf{H}},\mathbf{b}^{\star})}}\;,
		\label{eq:WaterFilingLikeSolutionDiag}%
	\end{equation}%
	where 
	\begin{equation*}%
		\psi(\boldsymbol{\mathsf{H}},\mathbf{b}^{\star})\coloneqq\min_{\mathbf{b}\in\mathcal{X}}\prod_{i=1}^n\varphi(|h_{ii}|B_i)
	\end{equation*}%
	with $\mathbf{b}\coloneqq(B_1,\dots,B_n)$ and $\varphi:\mathbb{R}_+\to\mathbb{R}_+$,
	\begin{equation}%
		%&\text{s.t. }      {\bf b} \in\mathcal{X} ,   \text{ where }   {\bf b}=[b_1,b_2,...,b_n],\\
		\varphi(x)\coloneqq\frac{1}{x^2}\parentheses*{\eu^{-x^2}-1 +\sqrt{\pi}x\bigl(1-2Q(\sqrt{2}x)\bigr)}\;,
		\label{eq:phi_function}%
	\end{equation}%
	and
	\begin{equation}%
		\barbelow{C}_{\mathsf{EPI}}(\mathcal{X},\boldsymbol{\mathsf{H}})=\frac{n}{2}\log\parentheses*{1+\mathrm{Vol}(\mathcal{X})^{\frac{2}{n}}\frac{\left|\prod_{i=1}^nh_{ii}\right|^{\frac{2}{n}}}{2\pi\eu}}\;.
		\label{eq:parrallelEPI}%
	\end{equation}%
\end{theorem}%
\begin{IEEEproof}%
	For some given values $B_i\in\mathbb{R}_+$, $i=1,\dots,n$, let the $i$-th component of $\X=(X_1,\dots,X_n)$ be independent and uniformly distributed over the interval $[-B_i,B_i]$. Thus, the expected value appearing in the bound of Theorem~\ref{thm:Jensen'sBound} can be written as
	\begin{equation}%
		\E\biggl[\eu^{-\frac{\|\boldsymbol{\mathsf{H}}(\X-\X')\|^2}{4}}\biggr]=\E\biggl[\eu^{-\frac{\sum_{i=1}^n h_{ii}^2(X_i-X'_i)^2}{4}}\biggr]=\E\Biggl[\prod_{i=1}^n\eu^{-\frac{h_{ii}^2(X_i-X'_i)^2}{4}}\Biggr]=\prod_{i=1}^n\E\biggl[\eu^{-\frac{h_{ii}^2(X_i-X'_i)^2}{4}}\biggr]\label{eq:expected_value_phi}\;.
	\end{equation}%
	Now, if $\mathbf{X}'$ is an independent copy of $\X$, it can be shown that the expected value at the right-hand side of \eqref{eq:expected_value_phi} is of the explicit form
	\begin{equation*}%
		\E\biggl[\eu^{-\frac{h_{ii}^2 (x_i-x'_i)^2}{4}}\biggr]=\varphi(|h_{ii}|B_i)\;
	\end{equation*}%
	with $\varphi$ as defined in \eqref{eq:phi_function}. Finally, optimizing over all $\mathbf{b}=(B_1,\dots,B_n)\in\mathcal{X}$ results in the bound \eqref{eq:WaterFilingLikeSolutionDiag}. The bound in \eqref{eq:parrallelEPI} follows by inserting $|\mathrm{det}(\boldsymbol{\mathsf{H}})|=\left|\prod_{i=1}^n h_{ii}\right|$ into \eqref{eq:AssumingInvertability}, which concludes the proof. 
\end{IEEEproof}%
\newsavebox{\smlmat}% Box to store smallmatrix content
\savebox{\smlmat}{$\boldsymbol{\mathsf{H}}=\left(\begin{smallmatrix}0.3 & 0 \\ 0 & 0.1\end{smallmatrix}\right)$}
\begin{figure}[t!]%
	\centering%
	\begin{tikzpicture}
	\pgfplotsset{every axis plot/.append style={very thick}}
	
	% Axis at [0.13 0.11 0.78 0.82]
	\begin{axis}[
	%axis on top,
	%scale only axis,
	width=0.51\textwidth,
	height=0.39\textwidth,
	xmin=0, xmax=30,
	ymin=0, ymax=15,
	xlabel={Amplitude Constraint, $A$, in dB},
	ylabel={Rate in $\text{bits}/\text{s}/\text{Hz}$},
	xmajorgrids,
	ymajorgrids,
	legend style={cells={anchor=west}, font=\footnotesize, at={(0.662,0.962)}}
	]
	
	\addplot [
	color=PU_darkgray,
	solid
	]coordinates{
		(0,0.0642381) (0.769231,0.0839232) (1.53846,0.103584) (2.30769,0.164235) (3.07692,0.237417) (3.84615,0.333201) (4.61538,0.45262) (5.38462,0.609249) (6.15385,0.804243) (6.92308,1.03565) (7.69231,1.3007) (8.46154,1.58691) (9.23077,1.89695) (10,2.22777) (10.7692,2.57301) (11.5385,2.93025) (12.3077,3.30867) (13.0769,3.70497) (13.8462,4.12325) (14.6154,4.55812) (15.3846,5.01312) (16.1538,5.48519) (16.9231,5.96382) (17.6923,6.45408) (18.4615,6.95076) (19.2308,7.44986) (20,7.95338) (20.7692,8.45929) (21.5385,8.96618) (22.3077,9.47357) (23.0769,9.98328) (23.8462,10.4926) (24.6154,11.0028) (25.3846,11.5129) (26.1538,12.0238) (26.9231,12.5346) (27.6923,13.0454) (28.4615,13.5563) (29.2308,14.0672) (30,14.5781)
	};
	\addlegendentry{Moment upper bound, $\bar{C}_{\mathsf{M}}$}
	
	\addplot [
	color=PU_darkorange,
	dash pattern=on 5pt off 3pt
	]coordinates{
		(0,0.135594) (0.769231,0.160981) (1.53846,0.190932) (2.30769,0.226194) (3.07692,0.267612) (3.84615,0.316127) (4.61538,0.372779) (5.38462,0.4387) (6.15385,0.515105) (6.92308,0.603274) (7.69231,0.704528) (8.46154,0.820203) (9.23077,0.951609) (10,1.09999) (10.7692,1.26649) (11.5385,1.4521) (12.3077,1.65764) (13.0769,1.88369) (13.8462,2.13062) (14.6154,2.39854) (15.3846,2.68729) (16.1538,2.99646) (16.9231,3.32543) (17.6923,3.67334) (18.4615,4.03918) (19.2308,4.42177) (20,4.81986) (20.7692,5.23213) (21.5385,5.65725) (22.3077,6.09389) (23.0769,6.5408) (23.8462,6.99676) (24.6154,7.46067) (25.3846,7.93149) (26.1538,8.40831) (26.9231,8.8903) (27.6923,9.37673) (28.4615,9.86696) (29.2308,10.3604) (30,10.8566)
	};
	\addlegendentry{Duality upper bound, $\bar{C}_{\mathsf{Dual},2}$}
	
	\addplot [
	color=PU_orange,
	dash pattern=on 1pt off 4pt on 4pt off 4pt
	]coordinates{
		(0,0) (9.23077,0) (10,0.0423632) (10.7692,0.197248) (11.5385,0.38442) (12.3077,0.603856) (13.0769,0.855128) (13.8462,1.13827) (14.6154,1.45361) (15.3846,1.80076) (16.1538,2.17766) (16.9231,2.58044) (17.6923,3.00434) (18.4615,3.4449) (19.2308,3.89857) (20,4.36267) (20.7692,4.83513) (21.5385,5.31435) (22.3077,5.79909) (23.0769,6.28833) (23.8462,6.78127) (24.6154,7.27727) (25.3846,7.77578) (26.1538,8.27639) (26.9231,8.77873) (27.6923,9.28251) (28.4615,9.78749) (29.2308,10.2935) (30,10.8003)
	};
	\addlegendentry{Jensen's lower bound, $\barbelow{C}_{\mathsf{Jensen}}$}
	
	\addplot [
	color=black,
	dotted
	]coordinates{
		(0,0.000633383) (0.769231,0.000902551) (1.53846,0.00128606) (2.30769,0.00183242) (3.07692,0.00261068) (3.84615,0.00371905) (4.61538,0.00529713) (5.38462,0.00754308) (6.15385,0.0107378) (6.92308,0.0152783) (7.69231,0.0217245) (8.46154,0.0308614) (9.23077,0.0437833) (10,0.0620006) (10.7692,0.087571) (11.5385,0.123246) (12.3077,0.172608) (13.0769,0.240158) (13.8462,0.331267) (14.6154,0.451904) (15.3846,0.608067) (16.1538,0.80493) (16.9231,1.04589) (17.6923,1.33183) (18.4615,1.66093) (19.2308,2.02905) (20,2.43062) (20.7692,2.85956) (21.5385,3.31008) (22.3077,3.77709) (23.0769,4.25642) (23.8462,4.74479) (24.6154,5.23972) (25.3846,5.73935) (26.1538,6.24234) (26.9231,6.74771) (27.6923,7.25476) (28.4615,7.76301) (29.2308,8.27209) (30,8.78177)
	};
	\addlegendentry{EPI lower bound, $\barbelow{C}_{\mathsf{EPI}}$}
	
	\end{axis}
	
	\end{tikzpicture}
	\vspace{-5pt}%
	\caption{Comparison of the upper and lower bounds of Theorems~\ref{thm:UpperBoundMaxEntropy}, \ref{thm:UpperBoundsParallel}, and \ref{thm:lowerBoundsParallel} evaluated for a $2\times 2$ MIMO system with per-antenna amplitude constraints $A_1=A_2=A$ (i.e., $\mathbf{a}=(A,A)$) and channel matrix~\usebox{\smlmat}.}
	\label{fig:Parallel}%
\end{figure}
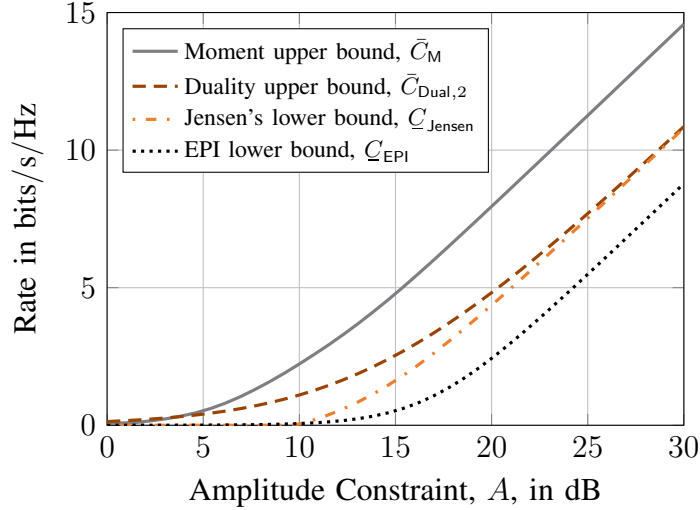%

In Fig.~\ref{fig:Parallel}, the upper bounds of Theorems~\ref{thm:UpperBoundMaxEntropy} and $\ref{thm:UpperBoundsParallel}$ and the lower bounds of Theorem~\ref{thm:lowerBoundsParallel} are depicted for a diagonal $2\times 2$ MIMO channel with per-antenna amplitude constraints. It turns out that the moment upper bound and the EPI lower bound perform well in the small amplitude regime while the duality upper bound and Jensen's inequality lower bound perform well in the high amplitude regime (note that the Jensen's inequality lower bound becomes strictly positive around $9\,\mathrm{dB}$).
%
%
%
%%%%%%%%%%%%%%%%%%%%%%%%%%%%%%%%%%%%%%%%%%%%%%%%%%%%%%%%%%%%%%%%%%%%%%%%%%%%%%%%%%%%%%%%%%%%%%%%%%%%%%%%%%%%%%%%%%%%%%%%%%%
\subsection{Gap to the Capacity}\label{sec:gap_results}%
Our first result bounds the gap between the capacity \eqref{eq:capacity_amp_constr} and the lower bound in \eqref{eq:AssumingInvertability}.
\begin{theorem}\label{thm:PackingEfficiecy}
	Let $\boldsymbol{\mathsf{H}}\in\mathbb{R}^{n \times n}$ be of full rank and
	\begin{equation*}%
		\rho(\mathcal{X},\boldsymbol{\mathsf{H}})\coloneqq\frac{\mathrm{Vol}\parentheses[\big]{\mathcal{B}_{\mathbf{0}}\left(r_{\mathsf{max}}(\boldsymbol{\mathsf{H}}\mathcal{X})\right)}}{\mathrm{Vol}(\boldsymbol{\mathsf{H}}\mathcal{X})}\;.
		\label{eq:packEfficiency}
	\end{equation*}%
	Then, 
	\begin{equation*}%
		C(\mathcal{X},\boldsymbol{\mathsf{H}})-\barbelow{C}_{\mathsf{EPI}}(\mathcal{X},\boldsymbol{\mathsf{H}})\le\frac{n}{2}\log\parentheses*{(\pi n)^{\frac{1}{n}}\rho(\mathcal{X},\boldsymbol{\mathsf{H}})^{\frac{2}{n}}}\;.
	\end{equation*}% 
\end{theorem}% 
\begin{IEEEproof}%
	For notational convenience let the volume of an $n$-dimensional ball of radius $r>0$ be denoted as 
	\begin{equation*}%
		V_n(r)\coloneqq\mathrm{Vol}\parentheses[\big]{\mathcal{B}_{\mathbf{0}}(r)}=V_n(1) r^n=\frac{\pi^{\frac{n}{2}}r^n}{\Gamma\parentheses*{\frac{n}{2}+1}}\;.
	\end{equation*}%
	
	Now, observe that by choosing $p=2$, the upper bound of Theorem~\ref{thm:UpperBoundMaxEntropy} can further be upper bounded as
	\begin{align*}
		\bar{C}_{\mathsf{M}}(\mathcal{X},\boldsymbol{\mathsf{H}})&\le n\log\parentheses*{\frac{k_{n,2}}{(2\pi\eu)^{\frac{1}{2}}}\,n^{\frac{1}{2}}\|\tilde{\mathbf{x}}+\Z\|_2}\\
		%& \stackrel{a)}{ \le} \inf_{p>1}  n \log \left( \frac{k_{n,p} }{ (2 \pi \eu)^{\frac{1}{2}}} \cdot n^{\frac{1}{p}} \cdot   \left( \frac{1}{n^{\frac{1}{p}}} \| \boldsymbol{\mathsf{H}} \bar{{\bf x}} \| + \|\Z\|_p \right) \right)\\
		&\stackrel{a)}{=}\frac{n}{2}\log\parentheses*{\frac{1}{n}\E\bigl[\|\tilde{\mathbf{x}}+\Z\|^2\bigr]}\\
		&\stackrel{b)}{=}\frac{n}{2}\log\parentheses*{1 + \frac{1}{n}\|\tilde{\mathbf{x}}\|^2}\;,
	\end{align*}%
	where $a)$ follows since $k_{n,2}=\sqrt{\frac{2\pi\eu}{n}}$ and $b)$ since $\E[\|\Z\|^2]=n$. Therefore, the gap between \eqref{eq:AssumingInvertability} and the moment upper bound of Theorem~\ref{thm:UpperBoundMaxEntropy} can be upper bounded as follows:
	\begin{align*}
		\bar{C}_{\mathsf{M}}(\mathcal{X},\boldsymbol{\mathsf{H}})-\barbelow{C}_{\mathsf{EPI}}(\mathcal{X},\boldsymbol{\mathsf{H}})&=\frac{n}{2}\log\parentheses*{\!\frac{1+\frac{1}{n}\|\tilde{\mathbf{x}}\|^2}{1+\frac{\mathrm{Vol}(\boldsymbol{\mathsf{H}}\mathcal{X})^{\frac{2}{n}}}{2\pi\eu}}\!}\\
			&\stackrel{a)}{=}\frac{n}{2}\log\parentheses*{\!\frac{1+\frac{1}{n}\parentheses*{\frac{V_n(\|\tilde{\mathbf{x}}\|)}{V_n(1)}}^{\frac{2}{n}}}{1+\frac{\mathrm{Vol}(\boldsymbol{\mathsf{H}}\mathcal{X})^{\frac{2}{n}}}{2\pi\eu}}\!}\\
			&=\frac{n}{2}\log\parentheses*{\!\frac{1+\frac{1}{n}\parentheses*{\frac{\rho(\mathcal{X},\boldsymbol{\mathsf{H}})\mathrm{Vol}(\boldsymbol{\mathsf{H}}\mathcal{X})}{V_n(1)}}^{\frac{2}{n}}}{1+\frac{\mathrm{Vol}(\boldsymbol{\mathsf{H}}\mathcal{X})^{\frac{2}{n}}}{2\pi\eu}}\!}\\
			%&=\frac{n}{2} \log\parentheses*{ n   \frac{  \rho(\mathcal{X},\boldsymbol{\mathsf{H}})^{\frac{2}{n}} \left( \frac{1}{V_n} \right)^{\frac{2}{n}}  \frac{1}{n} \mathrm{Vol}(\boldsymbol{\mathsf{H}}\mathcal{X})^{\frac{2}{n}} + 1}{1+\frac{\mathrm{Vol}(\boldsymbol{\mathsf{H}}\mathcal{X})^{\frac{2}{n}}}{2\pi\eu}}    \!}\\
			&\stackrel{b)}{\le}\frac{n}{2}\log\parentheses*{\!\frac{1}{n}2\pi\eu\left(\frac{\rho(\mathcal{X},\boldsymbol{\mathsf{H}})}{V_n(1)}\right)^{\frac{2}{n}}\!}\\
			%& = \frac{n}{2} \log\parentheses*{  2 \pi \eu  \,\rho(\mathcal{X},\boldsymbol{\mathsf{H}})^{\frac{2}{n}} \left( \frac{1}{V_n} \right)^{\frac{2}{n}}    \!}\\
			%& \stackrel{c)}{ \le} \frac{n}{2} \log\parentheses*{   \pi^{\frac{1}{n}}  \,\rho(\mathcal{X},\boldsymbol{\mathsf{H}})^{\frac{2}{n}}  n^{1+\frac{1}{n}}}\;.
			&\stackrel{c)}{\le}\frac{n}{2}\log\parentheses*{\!(\pi n)^{\frac{1}{n}}\rho(\mathcal{X},\boldsymbol{\mathsf{H}})^{\frac{2}{n}}\!}\;.
		\end{align*}
		Here, $a)$ is due to the fact that $\|\tilde{\mathbf{x}}\|$ is the radius of an $n$-dimensional ball, $b)$ follows from the inequality $\frac{1+cx}{1+x}\le c$ for $c \ge 1$ and $x\in\mathbb{R}_+$, and $c)$ follows from using Stirling's approximation to obtain $\left(\frac{1}{V_n(1)}\right)^{\frac{2}{n}}\le\frac{1}{2\eu\pi^{1-\frac{1}{n}}}n^{1+\frac{1}{n}}$.% This concludes the proof.
\end{IEEEproof}%

The term $\rho(\mathcal{X},\boldsymbol{\mathsf{H}})$ is referred to as the \emph{packing efficiency} of the set $\boldsymbol{\mathsf{H}}\mathcal{X}$. In the following proposition, we present the packing efficiencies for important special cases.
\begin{prop}\emph{(Packing Efficiencies)}%
	 \;Let $\boldsymbol{\mathsf{H}}\in\mathbb{R}^{n \times n}$ be of full rank, $A\in\mathbb{R}_+$, and $\mathbf{a}\coloneqq(A_1,\dots,A_n)\in\mathbb{R}_+^n$. Then,
	\begin{align}%
		\rho\parentheses[\big]{\mathcal{B}_{\mathbf{0}}(A),\boldsymbol{\mathsf{I}}_n}&=1\;, \label{eq:identityAndBall}\\
		\rho\parentheses[\big]{\mathcal{B}_{\mathbf{0}}(A),\boldsymbol{\mathsf{H}}}&=\frac{\|\boldsymbol{\mathsf{H}}\|^n}{|\mathrm{det}(\boldsymbol{\mathsf{H}})|}\;,\label{eq:HAndBall}\\
		\rho\parentheses[\big]{\mathrm{Box}(\mathbf{a}),\boldsymbol{\mathsf{I}}_n}&=\frac{\pi^{\frac{n}{2}}}{\Gamma\parentheses*{\frac{n}{2}+1}}\frac{\|\mathbf{a}\|^n}{\prod_{i=1}^nA_i}\;, \label{eq:IdentityAndCube}\\
		\rho\parentheses[\big]{\mathrm{Box}(\mathbf{a}),\boldsymbol{\mathsf{H}}}&\le\frac{\pi^{\frac{n}{2}}}{\Gamma\parentheses*{\frac{n}{2}+1}}\frac{\|\boldsymbol{\mathsf{H}}\|^n\|\mathbf{a}\|^n}{|\mathrm{det}(\boldsymbol{\mathsf{H}})|\prod_{i=1}^nA_i}\;.\label{eq:HAndCube}
	\end{align}%
\end{prop}%
\begin{IEEEproof}%
	The packing efficiency \eqref{eq:identityAndBall} follows immediately. Note that
	\begin{equation*}%
		r_{\mathsf{max}}\parentheses[\big]{\boldsymbol{\mathsf{H}}\mathcal{B}_{\mathbf{0}}(A)}=\max_{\mathbf{x}\in\mathcal{B}_{\mathbf{0}}(A)}\|\boldsymbol{\mathsf{H}}\mathbf{x}\|=\|\boldsymbol{\mathsf{H}}\|A\;.
	\end{equation*}%
	Thus, as $\boldsymbol{\mathsf{H}}$ is assumed to be invertible we have $\mathrm{Vol}(\boldsymbol{\mathsf{H}}\mathcal{B}_{\mathbf{0}}(A))=|\mathrm{det}(\boldsymbol{\mathsf{H}})|\mathrm{Vol}(\mathcal{B}_{\mathbf{0}}(A))$, which results in \eqref{eq:HAndBall}. To show \eqref{eq:IdentityAndCube}, observe that
	\begin{equation*}%
		\mathrm{Vol}\parentheses[\big]{\mathcal{B}_{\mathbf{0}}\parentheses[\big]{r_{\mathsf{max}}\parentheses[\big]{\boldsymbol{\mathsf{I}}_n\mathrm{Box}(\mathbf{a})}}}=\mathrm{Vol}\parentheses[\big]{\mathcal{B}_{\mathbf{0}}(\|\mathbf{a}\|)}=\frac{\pi^{\frac{n}{2}}}{\Gamma\parentheses*{\frac{n}{2}+1}}\|\mathbf{a}\|^n\;.
	\end{equation*}%
	The proof of \eqref{eq:IdentityAndCube} is concluded by observing that $\mathrm{Vol}(\boldsymbol{\mathsf{I}}_n\mathrm{Box}(\mathbf{a}))=\prod_{i=1}^nA_i$. Finally, observe that $\mathrm{Box}(\mathbf{a})\subset   \mathcal{B}_{\mathbf{0}}(\|\mathbf{a}\|)$ implies $r_{\mathsf{max}}(\boldsymbol{\mathsf{H}}\mathrm{Box}(\mathbf{a}))\le r_{\mathsf{max}}(\boldsymbol{\mathsf{H}}\mathcal{B}_{\mathbf{0}}(\|\mathbf{a}\|))$ so that
	\begin{equation*}%
		\rho\parentheses[\big]{\boldsymbol{\mathsf{H}},\mathrm{Box}(\mathbf{a})}\le\frac{\mathrm{Vol}\parentheses[\big]{\mathcal{B}_{\mathbf{0}}(\|\boldsymbol{\mathsf{H}}\|\|\mathbf{a}\|)}}{\mathrm{Vol}\parentheses[\big]{\boldsymbol{\mathsf{H}}\mathrm{Box}(\mathbf{a})}}=\frac{\pi^{\frac{n}{2}}}{\Gamma\parentheses*{\frac{n}{2}+1}}\frac{\|\boldsymbol{\mathsf{H}}\|^n\|\mathbf{a}\|^n}{|\mathrm{det}(\boldsymbol{\mathsf{H}})|\prod_{i=1}^n A_i}\;,
	\end{equation*}%
	which is the bound in \eqref{eq:HAndCube}.
\end{IEEEproof}%

\begin{figure}[!t]%
	\centering%
	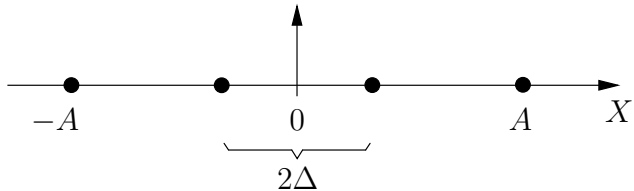%
	\caption{Example of a pulse-amplitude modulation constellation with $N=4$ points and amplitude constraint $A$ (i.e., $\pam(4,A)$), where $\Delta\coloneqq A/(N-1)$ denotes half the Euclidean distance between two adjacent constellation points. In case $N$ is odd, $0$ is a constellation point.}
	\label{fig:PAM_constellation}%
\end{figure}

We conclude this section by characterizing the gap to the capacity when $\boldsymbol{\mathsf{H}}$ is diagonal and the channel input space is the Cartesian product of $n$ PAM constellations. In this context, $\pam(N,A)$ refers to the set of $N\in\mathbb{N}$ equidistant PAM-constellation points with amplitude constraint $A\in\mathbb{R}_+$ (see Fig.~\ref{fig:PAM_constellation} for an illustration), whereas $X\sim\pam(N,A)$ means that $X$ is uniformly distributed over $\pam(N,A)$ \cite{dytsoISIT2017ozarow}. 
\begin{theorem}\label{thm:PAMparallel}% 
	Let $\boldsymbol{\mathsf{H}}=\mathrm{diag}(h_{11},\dots,h_{nn})\in\mathbb{R}^{n\times n}$ be fixed and $\X=(X_1,\dots,X_n)$. Then, if $X_i\sim\pam(N_i,A_i)$, $i=1,\dots,n$, for some given $\mathbf{a}=(A_1,\dots,A_n)\in\mathbb{R}_+^n$, it holds that
	\begin{equation}
		\bar{C}_{\mathsf{Dual},2}(\mathrm{Box}(\mathbf{a}),\boldsymbol{\mathsf{H}})-\barbelow{C}_{\mathsf{OW}}(\mathrm{Box}(\mathbf{a}),\boldsymbol{\mathsf{H}})\le  c\cdot n\,\textup{bits}\;,
		\label{eq:gapForOWParallel}
	\end{equation}%
	where $N_i\coloneqq\left\lfloor 1+\frac{2A_i |h_{ii}|}{\sqrt{2\pi\eu}}\right\rfloor$ and  
	\begin{equation*}%
		c\coloneqq\log(2)+\frac{1}{2} \log \parentheses*{\frac{\pi\eu}{6}}+\frac{1}{2}\log\parentheses*{1+\frac{6}{\pi\eu}}\approx 1.64\;.
	\end{equation*}%
	Moreover, if $X_i\sim\pam(N_i,A)$, $i=1,\dots,n$, for some given $A\in\mathbb{R}_+$, it holds that
	\begin{equation}%
		\bar{C}_{\mathsf{Dual},2}(\mathcal{B}_{\mathbf{0}}(A),\boldsymbol{\mathsf{H}})-\barbelow{C}_{\mathsf{OW}}(\mathcal{B}_{\mathbf{0}}(A),\boldsymbol{\mathsf{H}})\le c\cdot n\,\textup{bits}\;,
		\label{eq:gapOWParBall}
	\end{equation}%
	where $N_i\coloneqq\left\lfloor 1+\frac{2A |h_{ii}|}{\sqrt{n}\sqrt{2\pi\eu}}\right\rfloor$.%  and where 
%	\begin{align}
%	c=\log(2)+\frac{1}{2} \log \parentheses*{ \frac{\pi \eu }{6}}+\frac{1}{2} \log \parentheses*{ 1+\frac{6}{  \pi \eu} } \approx 1.64.
%	\end{align}
\end{theorem}%
\begin{IEEEproof}%
	%Due to space constraints, the proof is deferred to the extended version of this paper \cite{dytso2017amplitude_longer}. 
	Since the channel matrix is diagonal, letting the channel input $\X$ be such that its elements $X_i$, $i=1,\dots,n$, are independent we have that 
	\begin{equation*}%
		I(\X;\boldsymbol{\mathsf{H}}\X +\Z)=\sum_{i=1}^nI(X_i;h_{ii}X_i+Z_i)\;. 
	\end{equation*}%
	
	Let $X_i\sim\pam(N_i,A_i)$ with $N_i\coloneqq\left\lfloor 1+\frac{2A_i |h_{ii}|}{\sqrt{2\pi\eu}}\right\rfloor$ and observe that half the Euclidean distance between any pair of adjacent points in $\pam(N_i,A_i)$ is equal to $\Delta_i\coloneqq A_i/(N_i-1)$ (see Fig.~\ref{fig:PAM_constellation}), $i=1,\dots,n$. In order to lower bound the mutual information $I(X_i; h_{ii} X_i+ Z_i)$, we use the bound of Theorem~\ref{prop:OWimproved} for $p=2$ and $n_t=1$. Thus, for some continuous random variable $U$ that is uniformly distributed over the interval $[-\Delta_i,\Delta_i)$ and independent of $X_i$ we have that  
	%\begin{align*}
	%X_i \sim \pam(N_i,P_i), \, N_i= \left\lfloor 1+\frac{2a_i |h_{ii}|}{\sqrt{2 \pi \eu}} \right\rfloor,\\
	%U \sim \mathsf{Unif} \left[-\frac{d_{\min}(X_i)}{2},\frac{d_{\min}(X_i)}{2}\right].
	%\end{align*}
	%Moreover, note that the amplitude constraint can be related to the second moment as follows
	%\begin{align}
	%2a_i&= (N_i-1) d_{\min}(X_i)=(N_i-1) \sqrt{\frac{12 P_i}{N_i^2-1}} \notag\\
	%&= \sqrt{\frac{N_i-1}{N_i+1}} \sqrt{12 P_i} . \label{eq:RelationBetweenPowerANdAmplitude}
	%\end{align}
	\begin{equation}%
		I(X_i;h_{ii}X_i+Z_i)\ge H(X_i)-\frac{1}{2}\log\parentheses*{\frac{\pi\eu}{6}}-\frac{1}{2}\log\parentheses*{\frac{\E\bigl[(U +X_i-g(Y_i))^2\bigr]}{\E[U^2]}}\;. \label{eq:OWboundDiagonal}
	\end{equation}%
	Now, note that the entropy term in \eqref{eq:OWboundDiagonal} can be lower bounded as
	\begin{equation}%
		H(X_i)=\log\parentheses*{\left\lfloor 1+\frac{2A_i|h_{ii}|}{\sqrt{2\pi\eu}}\right\rfloor}\ge\log\parentheses*{1+\frac{2A_i|h_{ii}|}{\sqrt{2\pi\eu}}}+\log(2)\;,  \label{eq:entropyLowerBound}
	\end{equation}%
	where we have used that $\lfloor x\rfloor\ge \frac{x}{2}$ for every $x\ge 1$. On the other hand, the last term in \eqref{eq:OWboundDiagonal} can be upper bounded by upper bounding its argument as follows:	
	\begin{align}%
		\frac{\E\bigl[(U +X_i-g(Y_i))^2\bigr]}{\E[U^2]}&\stackrel{a)}{=}1+\frac{3\E\bigl[(X_i-g(Y_i))^2\bigr]}{\Delta_i^2} \notag\\
		&\stackrel{b)}{\le}1+\frac{3\E[Z_i^2](N_i-1)^2}{A_i^2|h_{ii}|^2}\notag\\
		&=1+\frac{3(N_i-1)^2}{A_i^2|h_{ii}|^2}\notag\\
		&\stackrel{c)}{\le}1+\frac{3\left(\frac{2A_i|h_{ii}|}{\sqrt{2\pi\eu}}\right)^2}{A_i^2|h_{ii}|^2}\notag\\
		&=1+\frac{6}{\pi\eu}\;.\label{eq:gapOWparallel}
	\end{align}%
	Here, $a)$ follows from using that $X_i$ and $U$ are independent and $\E[U^2]=\frac{\Delta_i^2}{3}$, $b)$ from using the estimator $g(Y_i)=\frac{1}{h_{ii}} Y_i$, and $c)$ from $N_i=\left\lfloor 1+\frac{2A_i|h_{ii}|}{\sqrt{2\pi\eu}}\right\rfloor\le 1+\frac{2A_i|h_{ii}|}{\sqrt{2\pi\eu}}$. Combining \eqref{eq:OWboundDiagonal}, \eqref{eq:entropyLowerBound}, and \eqref{eq:gapOWparallel} results in the gap \eqref{eq:gapForOWParallel}.
	 
	The proof of the gap in \eqref{eq:gapOWParBall} follows along similar lines, which concludes the proof. 
\end{IEEEproof}%
%
%
%
%%%%%%%%%%%%%%%%%%%%%%%%%%%%%%%%%%%%%%%%%%%%%%%%%%%%%%%%%%%%%%%%%%%%%%%%%%%%%%%%%%%%%%%%%%%%%%%%%%%%%%%%%%%%%%%%%%%%%%%%%%%
%%%%%%%%%%%%%%%%%%%%%%%%%%%%%%%%%%%%%%%%%%%%%%%%%%%%%%%%%%%%%%%%%%%%%%%%%%%%%%%%%%%%%%%%%%%%%%%%%%%%%%%%%%%%%%%%%%%%%%%%%%%
%	Arbitrary Channel Matrices
%%%%%%%%%%%%%%%%%%%%%%%%%%%%%%%%%%%%%%%%%%%%%%%%%%%%%%%%%%%%%%%%%%%%%%%%%%%%%%%%%%%%%%%%%%%%%%%%%%%%%%%%%%%%%%%%%%%%%%%%%%%
%%%%%%%%%%%%%%%%%%%%%%%%%%%%%%%%%%%%%%%%%%%%%%%%%%%%%%%%%%%%%%%%%%%%%%%%%%%%%%%%%%%%%%%%%%%%%%%%%%%%%%%%%%%%%%%%%%%%%%%%%%%
\section{Arbitrary Channel Matrices}\label{sec:SingularValuesDecom}%
For a MIMO channel with an arbitrary channel matrix and an average power constraint, the capacity is achieved by a singular value decomposition (SVD) of the channel matrix (i.e., $\boldsymbol{\mathsf{H}}=\boldsymbol{\mathsf{U}}\boldsymbol{\mathsf{\Lambda}}\boldsymbol{\mathsf{V}}^{T}$) and considering the equivalent channel model
\begin{equation*}%
	\tilde{\Y}=\boldsymbol{\mathsf{\Lambda}}\tilde{\X}+\tilde{\Z}\;, 
\end{equation*}%
where $\tilde{\Y}\coloneqq\boldsymbol{\mathsf{U}}^T\Y$, $\tilde{\X}\coloneqq\boldsymbol{\mathsf{V}}^{T}\X$, and $\tilde{\Z}\coloneqq\boldsymbol{\mathsf{U}}^T\Z$, respectively. %As a matter of fact, the orthogonal precoding of $\X$ by $\boldsymbol{\mathsf{V}}^T$ does not affect the average power constraint.%: $\E[\|\boldsymbol{\mathsf{V}}^{T}\X\|^2]=\E[\X^T\boldsymbol{\mathsf{V}}^T\boldsymbol{\mathsf{V}}\X]=\E[\X^T\X]=\E[\|\X\|^2]$.

To provide lower bounds for channels with amplitude constraints and SVD precoding, we need the following lemma.
\begin{lem}\label{lem:existanceOfUnif}% 
	For any given orthogonal matrix $\boldsymbol{\mathsf{V}}\in\mathbb{R}^{n_t\times n_t}$ and constraint vector $\mathbf{a}=(A_1,\dots,A_{n_t})\in\mathbb{R}_+^{n_t}$ there exists a distribution $F_{\X}$ of $\X$ such that $\tilde{\X}=\boldsymbol{\mathsf{V}}^{T}\X$ is uniformly distributed over $\mathrm{Box}(\mathbf{a})$. Moreover, the components $\tilde{X}_1,\dots,\tilde{X}_{n_t}$ of $\tilde{\X}$ are mutually independent with $\tilde{X}_i$ uniformly distributed over $[-A_i,A_i]$, $i=1,\dots,n_t$.
\end{lem}%
\begin{IEEEproof}%
	Suppose that $\tilde{\X}$ is uniformly distributed over $\mathrm{Box}(\mathbf{a})$; that is, the density of $\tilde{\X}$ is of the form
	\begin{equation*}%
		f_{\tilde{\X}}(\tilde{\mathbf{x}})=\frac{1}{\mathrm{Vol}\parentheses[\big]{\mathrm{Box}(\mathbf{a})}}\;,\;\tilde{\mathbf{x}}\in\mathrm{Box}(\mathbf{a})\;. 
	\end{equation*}%
	Since $\boldsymbol{\mathsf{V}}$ is orthogonal, we have $\boldsymbol{\mathsf{V}}\tilde{\X}=\X$ and by the change of variable theorem for $\mathbf{x}\in\boldsymbol{\mathsf{V}}\mathrm{Box}(\mathbf{a})$
	\begin{align*}%
		f_{\X}(\mathbf{x})=\frac{1}{|\mathrm{det}(\boldsymbol{\mathsf{V}})|}f_{\tilde{\X}}(\boldsymbol{\mathsf{V}}^T\mathbf{x})=\frac{1}{|\mathrm{det}(\boldsymbol{\mathsf{V}})|\mathrm{Vol}\parentheses[\big]{\mathrm{Box}(\mathbf{a})}}=\frac{1}{\mathrm{Vol}\parentheses[\big]{\mathrm{Box}(\mathbf{a})}}\;
	\end{align*}%
	Therefore, such a distribution of $\X$ exists. 
\end{IEEEproof}% 
%However, for the amplitude constrain case the situation is very different and given that ${\bf x} \in\mathcal{X}$, it does not necessarily imply that  $\V^t {\bf x} \in\mathcal{X}$ for an arbitrary set $\mathsf{S}$.  In fact this is only possible if the set $\mathsf{S}$ a ball centered at the origin. 
%
%
%
%However, it is important to asses the performance of the singular value precoding for the amplitude constraint case.  
%
%We let  $\mathsf{S}_{\text{rot}}= \V^t \cdot\mathcal{X}$ and let the set of feasible inputs be defined by
%\begin{align}
%\mathsf{S}_{\text{new}}=\mathcal{X}_{\text{rot}}  \cap \mathcal{X}.
%\end{align}
%
\begin{theorem}\label{thm:svdAchieve}\emph{(Lower Bounds with SVD Precoding})%
	\;Let $\boldsymbol{\mathsf{H}}\in\mathbb{R}^{n_r\times n_t}$ be fixed, $n_{\mathsf{min}}\coloneqq\min(n_r,n_t)$, and $\mathcal{X}=\mathrm{Box}(\mathbf{a})$ for some $\mathbf{a}=(A_1,\dots,A_{n_t})\in\mathbb{R}^{n_t}_+$. Furthermore, let $\sigma_i$, $i=1,\dots,n_{\mathsf{min}}$, be the $i$-th singular value of $\boldsymbol{\mathsf{H}}$. Then,
	\begin{equation}%
		\barbelow{C}_{\mathsf{Jensen}}(\mathrm{Box}(\mathbf{a}),\boldsymbol{\mathsf{H}})=\log^+\parentheses*{\left(\frac{2}{\eu}\right)^{\frac{n_{\mathsf{min}}}{2}}\frac{1}{\psi(\boldsymbol{\mathsf{H}},\mathbf{b}^{\star})}}
		\label{eq:WaterFilingLikeSolutionSVD}
	\end{equation}%
	and
	\begin{equation}%
		\barbelow{C}_{\mathsf{EPI}}(\mathrm{Box}(\mathbf{a}),\boldsymbol{\mathsf{H}})=\frac{n_{\mathsf{min}}}{2}\log\parentheses*{1+\frac{\left|\prod_{i=1}^{n_{\mathsf{min}}}A_i\sigma_{i}\right|^{\frac{2}{n_{\mathsf{min}}}}}{2\pi\eu}}\;,
		\label{eq:EPIsvd}%
	\end{equation}%
	where 
	\begin{equation*}%
		\psi(\boldsymbol{\mathsf{H}},\mathbf{b}^{\star})\coloneqq\min_{\mathbf{b}\in\mathrm{Box}(\mathbf{a})}\prod_{i=1}^{n_{\mathsf{min}}}\varphi(\sigma_iB_i)
	\end{equation*}%
	with $\mathbf{b}\coloneqq(B_1,\dots,B_{n_t})$ and $\varphi$ as defined in \eqref{eq:phi_function}.
%	\begin{equation}%
%		%&\text{s.t. }      {\bf b} \in\mathcal{X} ,   \text{ where }   {\bf b}=[b_1,b_2,...,b_n],\\
%		\varphi(x)\coloneqq\frac{1}{x^2}\parentheses*{\eu^{-x^2}-1 +\sqrt{\pi}x\bigl(1-2Q(\sqrt{2}x)\bigr)}\;.
%		\label{eq:phi_function2}%
%	\end{equation}%
	%\begin{equation}%
	%	\barbelow{C}_{\mathsf{EPI}}(\mathrm{Box}(\mathbf{a}),\boldsymbol{\mathsf{H}})=\frac{n_{\mathsf{min}}}{2}\log\parentheses*{1+\frac{\left|\prod_{i=1}^{n_{\mathsf{min}}}A_i\sigma_{i}\right|^{\frac{2}{n_{\mathsf{min}}}}}{2\pi\eu}}\;. 
	%	\label{eq:EPIsvd}%
	%\end{equation}%
	%and where $\sigma_i$ is the $i$-th singular value of $\boldsymbol{\mathsf{H}}$. 
\end{theorem}%
\begin{IEEEproof}%
	Performing the SVD, the expected value in Theorem~\ref{thm:Jensen'sBound} can be written as
	\begin{equation*}
		\E\biggl[\eu^{-\frac{\|\boldsymbol{\mathsf{H}}(\X-\X')\|^2}{4}}\biggr]=\E\biggl[\eu^{-\frac{\|\boldsymbol{\mathsf{U\Lambda V}}^{T}(\X-\X')\|^2}{4}}\biggr]=\E\biggl[\eu^{-\frac{\|\boldsymbol{\mathsf{\Lambda V}}^T(\X-\X')\|^2}{4}}\biggr]=\E\biggl[\eu^{-\frac{\|\boldsymbol{\mathsf{\Lambda}}(\tilde{\X}-\tilde{\X}')\|^2}{4}}\biggr]\;. 
	\end{equation*}%
	By Lemma~\ref{lem:existanceOfUnif} there exists a distribution $F_{\X}$ such that the components of $\tilde{\X}$ are independent and uniformly distributed. Since $\boldsymbol{\mathsf{\Lambda}}$ is a diagonal matrix, we can use Theorem~\ref{thm:lowerBoundsParallel} to arrive at \eqref{eq:WaterFilingLikeSolutionSVD}.
	
	Note that by Lemma~\ref{lem:existanceOfUnif} there exists a distribution on $\X$ such that $\tilde{\X}$ is uniform over $\mathrm{Box}(\mathbf{a})\subset \mathbb{R}^{n_t}$ and $\boldsymbol{\mathsf{\Lambda}}\tilde{\X}$ is uniform over $\boldsymbol{\mathsf{\Lambda}}\mathrm{Box}(\mathbf{a})\subset\mathbb{R}^{n_{\mathsf{min}}}$, respectively. Therefore, by the EPI lower bound given in \eqref{eq:lowerbound:EPI} we obtain
	\begin{align*}
		\barbelow{C}_{\mathsf{EPI}}(\mathrm{Box}(\mathbf{a}),\boldsymbol{\mathsf{H}})&=\frac{n_{\mathsf{min}}}{2}\log\parentheses*{1+\frac{2^{\frac{2}{n_{\mathsf{min}}}h(\boldsymbol{\mathsf{\Lambda}}\tilde{\X})}}{2\pi\eu}}\\
		&=\frac{n_{\mathsf{min}}}{2}\log\parentheses*{1+\frac{\mathrm{Vol}\parentheses[\big]{\boldsymbol{\mathsf{\Lambda}}\mathrm{Box}(\mathbf{a})}^{\frac{2}{n_{\mathsf{min}}}}}{2\pi\eu}}\\
		&=\frac{n_{\mathsf{min}}}{2}\log\parentheses*{1+\frac{\left(\prod_{i=1}^{n_{\mathsf{min}}}A_i\right)^{\frac{2}{n_{\mathsf{min}}}}\left|\prod_{i=1}^{n_{\mathsf{min}}}\sigma_{i}\right| ^{\frac{2}{n_{\mathsf{min}}}}}{2\pi\eu}}\;,
	\end{align*}%
	which is exactly the expression in \eqref{eq:EPIsvd}. This concludes the proof. 
\end{IEEEproof}%
\begin{remark}%
	Notice that choosing the optimal $\mathbf{b}$ for the lower bound \eqref{eq:WaterFilingLikeSolutionSVD} is an \emph{amplitude allocation problem}, which is reminiscent of waterfilling in the average power constraint case. It would be interesting to study whether the bound in \eqref{eq:WaterFilingLikeSolutionSVD} is connected to what is called \emph{mercury waterfilling} in \cite{OptimalPowerAllocationOfParChan,perez2010mimo}.
\end{remark}%
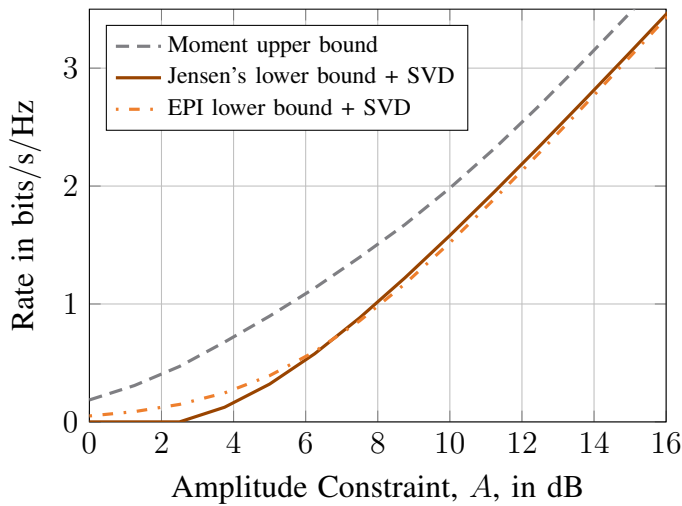
\begin{figure}%
	\centering%
	\begin{tikzpicture}
	\pgfplotsset{every axis plot/.append style={very thick}}
	
	% Axis at [0.13 0.11 0.78 0.82]
	\begin{axis}[
	%axis on top,
	%scale only axis,
	width=0.51\textwidth,
	height=0.39\textwidth,
	xmin=0, xmax=16,
	ymin=0, ymax=3.5,
	xlabel={Amplitude Constraint, $A$, in dB},
	ylabel={Rate in $\text{bits}/\text{s}/\text{Hz}$},
	xmajorgrids,
	ymajorgrids,
	legend style={cells={anchor=west}, font=\footnotesize, at={(0.657,0.962)}}
	]
	
	\addplot [
	color=PU_darkgray,
	dash pattern=on 5pt off 3pt
	]coordinates{
		(0,0.186189) (1.25,0.308559) (2.5,0.470477) (3.75,0.677031) (5,0.897226) (6.25,1.13505) (7.5,1.39669) (8.75,1.67248) (10,1.98248) (11.25,2.32201) (12.5,2.68789) (13.75,3.0729) (15,3.46977) (16.25,3.87451)
	};
	\addlegendentry{Moment upper bound}
	
	\addplot [
	color=PU_darkorange,
	solid
	]coordinates{
		(0,0) (2.5,0) (3.75,0.124329) (5,0.320708) (6.25,0.577979) (7.5,0.883145) (8.75,1.22093) (10,1.5807) (11.25,1.9557) (12.5,2.34149) (13.75,2.73503) (15,3.1342) (16.25,3.53752)
	};
	\addlegendentry{Jensen's lower bound + SVD}
	
	\addplot [
	color=PU_orange,
	dash pattern=on 1pt off 4pt on 4pt off 4pt
	]coordinates{
		(0,0.0502955) (1.25,0.0871484) (2.5,0.148365) (3.75,0.24584) (5,0.392095) (6.25,0.595826) (7.5,0.857631) (8.75,1.16968) (10,1.51968) (11.25,1.89546) (12.5,2.28752) (13.75,2.6894) (15,3.09701) (16.25,3.50793)
	};
	\addlegendentry{EPI lower bound + SVD}
	
	\end{axis}
	
	\end{tikzpicture}
	\vspace{-5pt}%
	\caption{Comparison of the upper bound in Theorem~\ref{thm:EntropyMax} to the lower bounds of Theorem~\ref{thm:svdAchieve} for a $3\times 1$ MIMO system with amplitude constraints $A_1=A_2=A_3=A$ (i.e., $\mathbf{a}=(A,A,A)$) and channel matrix $\boldsymbol{\mathsf{H}}=(0.6557,0.0357,0.8491)$.}
	\label{fig:charMiso}%
\end{figure}%

In Fig.~\ref{fig:charMiso}, the lower bounds of Theorem~\ref{thm:svdAchieve} are compared to the moment upper bound of Theorem~\ref{thm:EntropyMax} for the special case of a $3\times 1$ MIMO channel. Similarly to the example presented in Fig.~\ref{fig:Parallel}, the EPI lower bound performs well in the low amplitude regime while Jensen's inequality lower bound performs well in the high amplitude regime. 

We conclude this section by showing that for an arbitrary channel input space $\mathcal{X}$, in the large amplitude regime the capacity pre-log is given by $\min(n_r,n_t)$.
\begin{theorem}% 
	Let $\mathcal{X}$ be arbitrary and $\boldsymbol{\mathsf{H}}\in\mathbb{R}^{n_r\times n_t}$ fixed. Then,
	\begin{equation*}%
		\lim_{r_{\mathsf{min}}(\mathcal{X})\to\infty}\frac{C(\mathcal{X},\boldsymbol{\mathsf{H}})}{\log\parentheses*{1+\frac{2r_{\mathsf{min}}(\mathcal{X})}{\sqrt{2\pi\eu}}}}=\min(n_r,n_t)\;.
		\label{eq:HIGHsnrRegime}%
	\end{equation*}% 
\end{theorem}%
\begin{IEEEproof}%
	Notice that there always exists $\mathbf{a}\in\mathbb{R}_+^{n_t}$ and $c\in\mathbb{R}_+$ such that $\mathrm{Box}(\mathbf{a})\subseteq\mathcal{X}\subset c\mathrm{Box}(\mathbf{a})$. Thus, without loss generality we can consider $\mathcal{X}=\mathrm{Box}(\mathbf{a})$, $\mathbf{a}=(A,\dots,A)$, for sufficiently large $A\in\mathbb{R}_+$. To prove the result we therefore start with enlarging the constraint set of the bound in \eqref{eq:ProjecLikeBound}:
	\begin{align*}
		\mathrm{Box}\parentheses[\big]{\boldsymbol{\mathsf{H}}\mathrm{Box}(\mathbf{a})}&\subseteq\mathcal{B}_{\mathbf{0}}\parentheses[\big]{r_{\mathsf{max}}\bigl(\boldsymbol{\mathsf{H}}\mathrm{Box}(\mathbf{a})\bigr)}\\
		&\subseteq\mathcal{B}_{\mathbf{0}}\parentheses[\big]{r_{\mathsf{max}}\bigl(\boldsymbol{\mathsf{H}}\mathcal{B}_{\mathbf{0}}(\sqrt{n_t}A)\bigr)}\\
		&=\mathcal{B}_{\mathbf{0}}\parentheses[\big]{r_{\mathsf{max}}\bigl(\boldsymbol{\mathsf{U\Lambda V}}^T\mathcal{B}_{\mathbf{0}}(\sqrt{n_t}A)\bigr)}\\
		&=\mathcal{B}_{\mathbf{0}}\parentheses[\big]{r_{\mathsf{max}}\bigl(\boldsymbol{\mathsf{U\Lambda}}\mathcal{B}_{\mathbf{0}}(\sqrt{n_t}A)\bigr)}\\
		&=\mathcal{B}_{\mathbf{0}}\parentheses[\big]{r_{\mathsf{max}}\bigl(\boldsymbol{\mathsf{\Lambda}}\mathcal{B}_{\mathbf{0}}(\sqrt{n_t}A)\bigr)}\\
		&\subseteq\mathcal{B}_{\mathbf{0}}(r)\\
		&\subseteq\mathrm{Box}(\mathbf{a}')\;, 
	\end{align*}%
	where $r\coloneqq\sqrt{n_t}A\sqrt{\sum_{i=1}^{n_{\mathsf{min}}}\sigma^2_i}$ and $\mathbf{a}'\coloneqq\bigl(\frac{r}{\sqrt{n_{\mathsf{min}}}},\dots,\frac{r}{\sqrt{n_{\mathsf{min}}}}\bigr)\in\mathbb{R}_+^{n_{\mathsf{min}}}$. Therefore, by using the upper bound in \eqref{eq:ProjecLikeBound} it follows that
	\begin{equation*}%
		C(\mathrm{Box}(\mathbf{a}),\boldsymbol{\mathsf{H}})\le\sum_{i=1}^{n_r}\log\parentheses*{1+\frac{2 A_i}{\sqrt{2\pi\eu}}}\le n_{\mathsf{min}}\log\parentheses*{1+\frac{2}{\sqrt{2\pi\eu}}\frac{\sqrt{n_t}A\sqrt{\sum_{i=1}^{n_{\mathsf{min}}}\sigma^2_i}}{\sqrt{n_{\mathsf{min}}}}}\;.
	\end{equation*}%
	Moreover, 
	\begin{equation*}
		\lim_{A\to\infty}\frac{C(\mathrm{Box}(\mathbf{a}),\boldsymbol{\mathsf{H}})}{\log\parentheses*{1+\frac{2A}{\sqrt{2\pi\eu}}}}\le  n_{\mathsf{min}}\lim_{A\to\infty}\frac{\log\parentheses*{1+\frac{2}{\sqrt{2\pi\eu}}\frac{\sqrt{n_t}A\sqrt{\sum_{i=1}^{n_{\mathsf{min}}}\sigma^2_i}}{\sqrt{n_{\mathsf{min}}}}}}{\log \parentheses*{1+\frac{2A}{\sqrt{2\pi\eu}}}}=n_{\mathsf{min}}\;.
	\end{equation*}% 
	
	Next, using the EPI lower bound in \eqref{eq:EPIsvd}, we have that 
	\begin{equation*}%
		\lim_{A\to\infty}\frac{\barbelow{C}_{\mathsf{EPI}}(\mathrm{Box}(\mathbf{a}),\boldsymbol{\mathsf{\Lambda}})}{\log\parentheses*{1+\frac{2A}{\sqrt{2\pi\eu}}}}=n_{\mathsf{min}}\lim_{A\to\infty}\frac{\frac{1}{2}\log\parentheses*{1+\frac{A\left|\prod_{i=1}^{n_{\mathsf{min}}}\sigma_{i}\right|^{\frac{2}{n_{\mathsf{min}}}}}{2\pi\eu}}}{\log\parentheses*{1+\frac{2A}{\sqrt{2\pi\eu}}}}=n_{\mathsf{min}}\;.
	\end{equation*}%
	This concludes the proof. 
\end{IEEEproof}%
%
%
%
%%%%%%%%%%%%%%%%%%%%%%%%%%%%%%%%%%%%%%%%%%%%%%%%%%%%%%%%%%%%%%%%%%%%%%%%%%%%%%%%%%%%%%%%%%%%%%%%%%%%%%%%%%%%%%%%%%%%%%%%%%%
%%%%%%%%%%%%%%%%%%%%%%%%%%%%%%%%%%%%%%%%%%%%%%%%%%%%%%%%%%%%%%%%%%%%%%%%%%%%%%%%%%%%%%%%%%%%%%%%%%%%%%%%%%%%%%%%%%%%%%%%%%%
%	Conclusion
%%%%%%%%%%%%%%%%%%%%%%%%%%%%%%%%%%%%%%%%%%%%%%%%%%%%%%%%%%%%%%%%%%%%%%%%%%%%%%%%%%%%%%%%%%%%%%%%%%%%%%%%%%%%%%%%%%%%%%%%%%%
%%%%%%%%%%%%%%%%%%%%%%%%%%%%%%%%%%%%%%%%%%%%%%%%%%%%%%%%%%%%%%%%%%%%%%%%%%%%%%%%%%%%%%%%%%%%%%%%%%%%%%%%%%%%%%%%%%%%%%%%%%%
\section{Conclusion}\label{sec:conclusion}%
In this work, we have focused on studying the capacity of MIMO systems with bounded channel input spaces. Several new upper and lower bounds have been proposed and it has been shown that
the lower and upper bounds are tight in the high amplitude regime. An interesting direction for future work is to determine the exact scaling in the massive MIMO regime (i.e., $n_{\mathsf{min}}\to\infty$). 

%Another interesting future direction is to exploiting techniques of \cite{} to tighten our upper bounds  for the of massive MIMO regime and small amplitude constraint regime (i.e., $ r_{\mathsf{max}} \to0 $). 
Another interesting future direction is to study generalizations of our techniques to MIMO wireless optical channels \cite{mosercapacity}. 
%
%\subsection{Without SVD precoding}
%
%Without the SVD decomposition the following rate can be achieved. 
%\begin{theorem} \label{thm:WithoutSVD}   For any $\mathsf{S}$ and $\boldsymbol{\mathsf{H}}$
%\begin{align}
%R_{\text{Jensen}}(\mathsf{S},\boldsymbol{\mathsf{H}} ) &= - \log  \left( \frac{ \eu^{\frac{n_r}{2}}   }{ \sqrt{2}  }  g_B\right), \label{eq:WaterFilingLikeSolutionGeneral}
%\end{align}
%where 
%\begin{align}
%&g_B= \min_{ b_i: i=[1:n_t]}  \prod_{i=1}^{n_t}  f( \sigma_{\min} b_i),\\
%&\text{s.t. }     {\bf b} \in\mathcal{X} ,   \text{ where }   {\bf b}=[b_1,b_2,...,b_{n_t}],
%\end{align}
%where $\sigma_{\min} $ is the smallest singular value of $\boldsymbol{\mathsf{H}}$. 
%\end{theorem}
%\begin{IEEEproof}
%The proof follows by using the following lower bound 
%\begin{align}
%\| \boldsymbol{\mathsf{H}} (\X-\X_1)\| \ge  \sigma_{\min}  \| \X-\X_1\|,
%\end{align}
%in Theorem~\ref{thm:Jensen'sBound}.
%\end{IEEEproof} 
%
%Note, that  in Theorem~\ref{thm:WithoutSVD} the optimization is done over the original set $\mathsf{S}$ instead of $\mathsf{S}_{\text{new}}$. {\red Q: Can  this be better??? Need a simualtion.}
%
%
%
%
%

\end{document}